\documentclass[%
 reprint,
superscriptaddress,nofootinbib
]{revtex4-1}

\usepackage{graphicx}
\usepackage{dcolumn}
\usepackage{bm}
\usepackage{bm}
\usepackage{enumerate}
\usepackage{eurosym}
\usepackage{amsfonts}
\usepackage{amsmath}
\usepackage{amsopn}
\usepackage{amssymb}
\usepackage{graphicx}
\usepackage{graphics}
\usepackage[colorlinks=true, allcolors=blue]{hyperref}
\usepackage{color}
\usepackage{verbatim}
\usepackage{mathrsfs}
\usepackage{hyperref}
\usepackage{braket}

\begin{document}

\preprint{APS/123-QED}

\title{Synthetic plasmonic lattice formation through invariant frequency comb excitation in graphene structures}

\author{Zahra Jalali-Mola}
\affiliation{Independent Researcher, Personal Research Laboratory, West Ferdows Blvd., Tehran, 1483633987, Iran}
\email{sasgarnezhad93@gmail.com}

\author{Saeid Asgarnezhad-Zorgabad}
\affiliation{Independent Researcher, Personal Research Laboratory, West Ferdows Blvd., Tehran, 1483633987, Iran}

\date{\today}

\begin{abstract}
  Nonlinear surface-plasmon polaritons~(NSPPs) in nanophotonic waveguides excite with dissimilar temporal properties due to input field modifications and material characteristics, but they possess similar nonlinear spectral evolution. In this work, we uncover the origin of this similarity and establish that the spectral dynamics is an inherent property of the system that depends on the synthetic dimension and is beyond waveguide geometrical dimensionality. To this aim, we design an ultra-low loss nonlinear plasmonic waveguide, to establish the invariance of the surface plasmonic frequency combs~(FCs) and phase singularities for plasmonic peregrine waves and Akhmediev breather. By finely tuning the nonlinear coefficient of the interaction interface, we uncover the conservation conditions through this plasmonic system and employ the mean-value evolution of the quantum NSPP field commensurate with the Schr\"odinger equation to evaluate spectral dynamics of the plasmonic FCs~(PFCs). Through providing suppressed interface losses and modified nonlinearity as dual requirements for conservative conditions, we propose exciting PFCs as equally spaced invariant quantities of this plasmonic scheme and prove that the spectral dynamics of the NSPPs within the interaction interface yields the formation of plasmonic analog of the synthetic photonic lattice, which we termed \textit{synthetic plasmonic lattice}~(SPL).
\end{abstract}

\keywords{ultra-low loss graphene, plasmonic phase singularity, frequency combs, synthetic lattice.}
\maketitle
	
\section{Introduction}
Synthetic lattice~(SL)~\cite{FanReview2018} provides a platform for photonic structures to couple the integral degree of freedom of light such as orbital angular momentum and FC with geometrical dimensions of the waveguide to form higher-order synthetic space~\cite{Yuan:16,dutt2020higher}. This multidimensional property observes both theoretically and experimentally in various physical systems from photonics~\cite{Ozawa2019RMP,Kivshar2020APR} and cold atoms~\cite{Cooper2019RMP} to non-Hermitian systems~\cite{PhysRevApplied.14.064076} and topological circuits~\cite{imhof2018topolectrical}. SL also provides artificial gauge fields for a bosonic structure, which yields control over spectral and temporal behaviors of light and hence is valuable for topological lasing~\cite{Hararieaar4003,Bandreseaar4005}, breaking time-reversal symmetry~\cite{lumer2019light}, etc.

Recently, SL with the periodic-boundary condition is introduced to the reconstruction of the FCs~\cite{doi:10.1063/1.5144119}, to control the light manipulation in a nonlinear waveguide~\cite{wang2020multidimensional} and to induce a synthetic Hall effect for photons~\cite{doi:10.1063/5.0034291}. In previous investigations, these lattices are constructed as photonic structures with negligible dissipation and dispersion, whose internal degree of light acts as a synthetic dimension. Besides, plasmonic structures act as nanoscopic nonlinear waveguides that transport surface-plasmon polaritons~(SPPs) instead of photons. NSPP wave propagation is a well-explored topic within these media both in the presence and absence of gain~\cite{PhysRevA.99.051802,Asgarnezhad_Zorgabad_2020,Asgarnezhad-Zorgabad:20}. It is also well-known that the combination of nonlinear response and gain amplification can be exploited to excite and sustain nonlinear waves such as different classes of solitons. As these hybrid interfaces possess the same nonlinearity and dispersion, there must be similarities such as the internal degree of freedom between these nonlinear waves. As spatiotemporal profiles of these NSPPs are quite dissimilar, these hidden similarities should be beyond waveguide geometrical dimensionality and may be related to synthetic dimension.

Consequently, natural questions that may arise are whether we can propose an SL for plasmonic nanostructures, and what would be the consequence of this synthetic plasmonic lattice~(SPL)? Quite generally, constructing an SPL using an internal degree of freedom of SPPs has not yet been investigated and this concept should be a subject of potential applications from quantum nanophotonics~\cite{doi:10.1021/acsphotonics.0c01224} to ultrafast-nanoplasmonics~\cite{doi:10.1021/acs.nanolett.0c02452}. Note that our work is conceptually novel, as we introduce the concept of synthetic dimension to dissipative nanophotonic structures such as plasmonic waveguides, and also this work is methodologically novel, as we develop a framework based on quantum nonlinear averaging of SPP field, to uncover the similarities between various NSPPs, and to discover the invariants of a plasmonic scheme in a loss-compensated waveguide. Finally, we propose a general nonlinear plasmonic waveguide configuration that comprises a tunable nonlinear layer situated on top of a plasmonic scheme with vanishing loss, consequently, our work can be extended to other low-loss hybrid nanostructures. We justify SL formation within a nonlinear plasmonic structure in three steps, namely, (i) first, we elucidate the robust PFCs propagation, (ii) we uncover the conservative conditions, and (iii) we introduce loss suppression and careful nonlinearity modification as dual requirements to generate robust PFCs and to form SPL. 

\emph{Physical picture of our scheme}- In this work, we aim to uncover the mutual propagation properties between plasmonic peregrine wave and Akhmediev breather. The key result of this work is that in a plasmonic waveguide with suppressed loss and tunable nonlinear coefficient, PFCs act as an invariant across different types of nonlinear waves and we interpret them in terms of a synthetic dimension to construct an SPL. Consequently, our work introduces two novel concepts to nonlinear plasmonic nanostructures, namely, (i) unveiling PFCs as synthetic dimension, and (ii) exploiting this synthetic dimension to form an SPL. To elucidate these concepts, we design a nonlinear graphene nanostructure that fulfills conservation conditions and can be exploited to validate our results.

\begin{figure}[t]
\centering
\fbox{\includegraphics[width=1\linewidth]{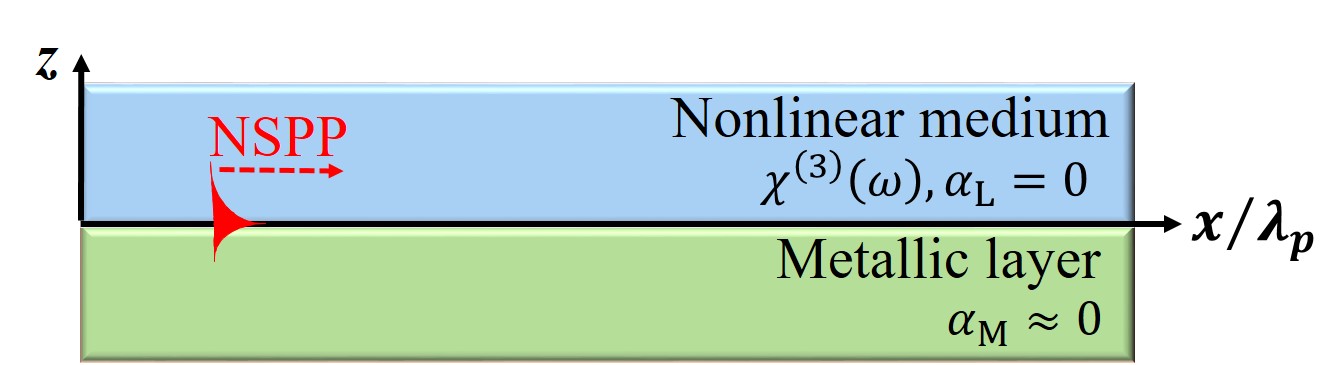}}
\caption{Nonlinear plasmonic waveguide configuration for exciting invariant PFC and constructing SPL. Our suggested scheme comprises a plasmonic scheme placed as the bottom layer and a nonlinear medium situated on top. This nonlinear waveguide should possesses tunable nonlinearity and vanishing loss. $\chi^{(3)}(\omega)$ is the nonlinear coefficient of the medium, $\alpha_\text{L}=0$ is its linear absorption coefficient, and we assume $\alpha_\text{M}:=\text{Im}[\bm{\varepsilon}(k,\omega)]$ as suppressed loss of the plasmonic layer.}
    \label{fig:Fig_One}
\end{figure}

\section{Model}\label{Sec:Model}
To uncover the invariant parameters of NSPPs, we suggest a nonlinear plasmonic nanostructure as it is shown in Fig.~\ref{fig:Fig_One}. This scheme is a general extension to nonlinear hybrid plasmonic configurations that comprise a nonlinear medium situated on top of a plasmonic layer. In order to achieve the robust PFCs, establish the conservation conditions, and construct the SPL, we require a nonlinear waveguide with vanishing loss and tunable nonlinearity. In what follows, we elucidate the key \textit{essential} components of each layer that validate the invariance of PFCs yielding formation of SPL.

\emph{Nonlinear material description}- Various materials possess optical nonlinearity that may serve as the upper layer of our nonlinear plasmonic scheme. However, we consider a nonlinear medium that simultaneously fulfills two criteria, namely, (i) possesses suppressed linear absorption, (ii) the nonlinear parameter related to this layer is tunable. This nonlinear medium supports different classes of nonlinear waves such as Akhmedive breather and peregrine wave and provides opportunities to excite and generate robust FCs, which are required to construct an SPL.

\emph{Plasmonic layer description}- Besides, our proposed plasmonic layer should possess ultra-low Ohmic loss for our wavelength of interest. Although various ultra-low loss schemes can serve as the bottom layer, our scheme is double-layer graphene that can be modeled as substrate-graphene-dielectric-graphene multi-layer, as it is indicated in Fig.~\ref{fig:Fig_One}. We introduce gain to the bottom graphene layer by trigger laser irradiation through a photo-inverted scheme and adjust a suitable laser power to suppress the Ohmic-loss of the waveguide through gain-loss competition. Our ultra-low loss plasmonic scheme is an extension to previous researches~\cite{PhysRevA.99.051802,Asgarnezhad_Zorgabad_2020,Asgarnezhad-Zorgabad:20}, for which we exploit the gain-induced loss compensation in double-layer graphene structure~(see \S~\textcolor{blue}{B} of the appendix) instead of employing negative-index metamaterial structure.

\emph{General description of excitation}- To sum up, we propose a nonlinear plasmonic waveguide that fulfills suppressed interface losses and modified nonlinearity as dual requirements for conservative conditions and supports different classes of NSPP waves such as plasmonic Akhemediev breather and peregrine waves. Robust PFCs excite as a consequence of these plasmonic fields that act as invariants of the system and can be interpreted in terms of synthetic dimension. The existence of this synthetic dimension would yield apparition of plasmonic counterparts of synthetic lattice that we termed as SPL.

\section{Approach}\label{Sec:Approach}
We present our quantitative approach towards robust NSPP propagation in three steps. First, we present the key parameters required to describe our nonlinear waveguide. Next, we describe the SPP field excitation in our scheme. Finally, we elucidate the Fourier evolution of NSPP fields.

\emph{Essential parameters related to scheme}- We quantitatively model our nonlinear waveguide in terms of the parameters related to the nonlinear medium and plasmonic scheme. We consider the nonlinear coefficient of the upper layer as $\chi^{(3)}(\omega)$, its linear absorption is $\alpha_\text{L}=0$ and we further assume that the suppressed linear absorption and enhanced nonlinear coefficient are simultaneously achievable through adjusting a control parameter of the system. Besides, our proposed plasmonic structure is double-layer graphene, the bottom layer possesses gain and the upper graphene is lossy. We consider the susceptibility of the lossy graphene as $\chi^{(\text{L})}(k,\omega)$~\cite{Jalaligraphene}, gain-assisted graphene as $\chi^{(\text{G})}(k,\omega)$~\cite{SHess2015} and evaluate the effective susceptibility of the coupled system as $\chi^{(\text{C})}(k,\omega)$\footnote{We present the detailed quantitative steps of derivation in \S~\textcolor{blue}{B} of appendix.}. Consequently, we achieve the dielectric function of the double-layer graphene as $\bm\varepsilon(k,\omega)=\bm1-\bm\chi^{(\text{C})}(k,\omega) \bm V(k)$; $\bm{V}(k)$ the coupling matrix between two layers. This double layer excites SPP for $\text{det}\{\bm{\varepsilon}(k,\omega)\}=0$~\cite{jalalidoublelayer}, and ultra low-loss SPP field propagates for $\text{Im}[\text{det}\{\bm\varepsilon(k,\omega)\}]=0$. The combination of the nonlinear material and this graphene nanostructure provides a nonlinear plasmonic waveguide that fulfills conservative conditions and hence is suitable to excite invariant PFCs.

\emph{Surface plasmon-field excitation}- The graphene structure-nonlinear medium interface hence excite stable SPP with reciprocal chromatic dispersion $\omega(\mathcal{K})=\omega(-\mathcal{K})$, with constant phase $\theta=\mathcal{K}(\omega)x_{l}-\omega t_{l}$ and with group velocity $v_\text{g}=[\partial\mathcal{K}/\partial\omega]^{-1}$. The group-velocity dispersion of the SPP field is $\mathcal{K}_{2}=\partial^{2}\mathcal{K}/\partial\omega^{2}$ and the self-focusing nonlinearity is $W>0$\footnote{$\mathcal{W}(\omega)\propto\chi^{(3)}(\omega)$ for a nonlinear material.}. We tuned these nonlinear coefficient through adjusting the control parameter. Next, we consider frequency grid as $f=\sqrt{(\mathcal{K}_{2}\delta\omega^{2})/(W\pi^{2})}$, temporal grid as $\tau_{0}\sim1/(\delta\omega)$, absorption coefficient as $\bar{\alpha}=\epsilon^{2}\text{Im}[\mathcal{K}(\omega)]+\text{Im}[k_\text{C}(\omega)]$; $\epsilon\ll1$ the perturbation parameter, and finally we normalize the probe pulse envelope as $u=[\Omega_\text{P}/\xi]\exp\{-\bar{\alpha}x\}$. This pulse is then stable in rotated time~($\tau=t-x/v_\text{g}$) for a few nonlinear $L_\text{NL}=1/(\xi^{2}W)$ and dispersion lengths $L_\text{D}=\tau_{0}^{2}/\mathcal{K}_{2}$. 

\emph{Spectral evolution of nonlinear SPP}- The evolution of the SPP field then depends on two nonlinear parameters, i.e. dispersion $\mathcal{K}_2(\omega)$, and nonlinearity $\mathcal{W}(\omega)$. Excited NSPPs propagate through an effective interface $S_\text{eff}$, within a characteristic time scale $t_\text{S}$ and evolution of these nonlinear fields would yield FCs generation. In a medium with higher-order dispersion $\mathcal{K}_{n}$, we define PFCs as discrete frequency components associated with compressed plasmonic pulse envelope that excite with a central frequency $\omega_\text{SPP}$, with frequency spacing $\delta$ that propagates through the interface as $\omega_n=\omega_\text{SPP}+\sum_n(D_n/n!)\delta^n$ for $D_n=\sum_{n={2,3,\ldots}}(\mathcal{K}_n/n!)\delta^n$. The $m$th PFCs has mode frequency $\omega_m$, amplitude $\tilde{A}_{m}(x,\omega_m)$ that would excite and propagate through the interface. Total energy of the FCs and the total number of excited plasmon modes are $E\propto\sum_m |\tilde{A}_{m}(x,\omega_m)|^2$ and $\mathcal{N}\propto\sum_m|\tilde{A}_{m}(x,\omega_m)|^2/(\omega_\text{p}+\omega_m)$, respectively. The total SPP field related to PFCs $\tilde{\Psi}:=\tilde{\Psi}(x,\tilde{\omega})=F(\bm{r})\tilde{A}(x,\tilde{\omega})\exp\{\text{i}\mathcal{K}(\tilde{\omega}) x\}$; $F(\bm{r}):=F(y,z)$ the plasmonic pulse envelope function, propagates through the interface whose dynamics described by nonlinear spectral evolution equation
\begin{equation}
	\text{i}\frac{\partial\tilde{\Psi}}{\partial x}=\mathcal{K}(\omega)\tilde{\Psi}+\sum_m\int\text{d}\tilde{\omega}\;\frac{\mathcal{W}(\tilde{\omega})}{2\pi}\tilde{\Psi}_{\tilde{\omega}}^{*}\tilde{A}_{\tilde{\omega}-\omega_m}\tilde{\Psi}_{\tilde{\omega}+\omega_m}^{*}.
	\label{Eq:Dynamics_SPP_Field}
\end{equation}
We note that the NSPP propagation within the nonlinear medium-graphene interface is similar to previous works~\cite{PhysRevA.99.051802,Asgarnezhad_Zorgabad_2020,Asgarnezhad-Zorgabad:20} as both systems possess group-velocity dispersion and self-phase modulation. However, this work differs from our previous works due to exploiting NSPP similarities to establish conservation conditions, introduce frequency comb as a synthetic dimension, and constructing SPL. 

\emph{Simulation parameters and scheme feasibility}- Now, we introduce the simulation parameters to justify invariant PFCs excitation. In our scheme stable NSPP with $\omega_\text{SPP}=1.57~\text{eV}$ propagates with group velocity dispersion $\mathcal{K}_{2}=(-4.42+0.4\text{i})\times10^{-12}\text{s}^{2}\cdot\text{cm}^{-1}$, and with nonlinear coefficient $W=(2.98+0.6\text{i})\times10^{-11}\text{s}^{2}\cdot\text{cm}^{-1}$, and the group velocity of the excited SPP is $v_\text{g}=2\times10^{4}\text{m}/\text{s}$. The normalized nonlinear parameter is $g_{\mathcal{K}_{2}}=1.01$ and frequency grid is $f=0.045~\text{s}^{-1}$. For gain-assisted graphene we have $\text{Im}[k_\text{G}]=0.07~\text{cm}^{-1}$ and for coupled system we have $\text{Im}[k_\text{C}]=-0.02~\text{cm}^{-1}$, hence $\bar{\alpha}=\text{Im}[K_\text{a}(\omega)+k_\text{C}(\omega)]=0.04\approx0$. Therefore, our proposed structure is the best fit to vanishing loss and tunable nonlinear parameter and hence is a suitable candidate for invariant PFCs excitation and SPL formation.

To simulate SPP propagation in this nonlinear plasmonic system, we assume (i) SPP waves propagate as far plasmonic fields, (ii) the plasmonic phases are constant~(i.e. $\mathcal{K}x-\omega t=\text{Const.}$), and (iii) employ mean field-averaging~\cite{Asgarnezhad-Zorgabad:20,Asgarnezhad_Zorgabad_2020,PhysRevA.98.013825} to consider the evanescent coupling effect. Finally, we note that our constructed synthetic lattice can also be achieved in other plasmonic schemes. To this aim, the interaction interface should satisfy the dual requirement of loss suppression and nonlinearity modulation. We note that our proposal is a general extension to other nonlinear plasmonic schemes, and the formation of SPL and invariant PFCs in any other configuration can also be investigated using our theory. We exploit our method for a plasmonic configuration comprises nonlinear atomic ensemble as a specific example and elucidate all detailed elucidations such as mechanisms required to excite and probe the system within this hybrid design, the details of the nonlinear medium, and all detailed calculations of graphene nanostructure in \S~\textcolor{blue}{A} of the appendix.  

\begin{figure}[t]
\centering
\fbox{\includegraphics[width = 1\linewidth]{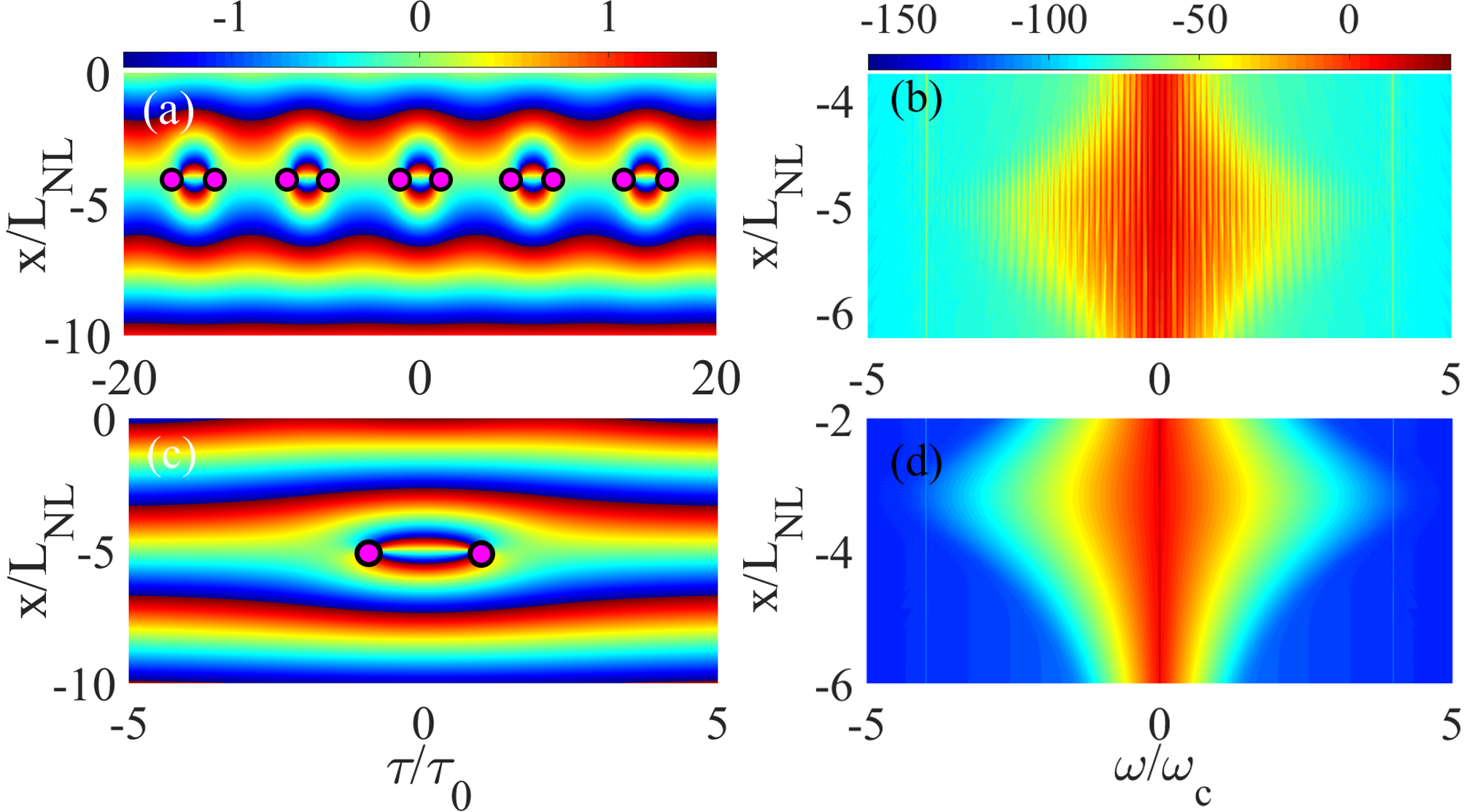}}
\caption{Panel (a) represents the phase dynamics of NSPPs through Akhmediev breather formation and panel (b) denotes corresponding spectral analysis. Panel (c) depicts the phase dynamics of NSPP through plasmonic peregrine wave and panel (d) represents its spectral evolution. For this figure $P_{0}=10~\mu\text{W}$, $\tau_{0}=10~\mu\text{s}$, $\delta\omega=1~\text{MHz}$, $u_\text{N}=0.08~\text{MHz}$. Panels (a) and (b) are plotted for modulation parameter $a=0.32$ and for panels (c) and (d) we choose $a=0.5$. Magenta dots in panels (a) and (c) represent the coordinates of PSs. Other parameters are given in the text.}
\label{Fig:Two}
\end{figure}

\section{Results}\label{Sec:Results}
We present the results of this paper in three sections: First, we investigate the temporal and spectral dynamics of the plasmonic peregrine and Akhmediev breather phases within the interaction interface in \S~\ref{Sec:Universality}. Next, in \S~\ref{Sec:Invariants} we evaluate the spatial-spectral evolution of the energy flux and number of plasmon modes to achieve the conservative parameters of the system. Finally, we map the robust spectral dynamics to a synthetic photonic lattice in \S~\ref{Sec:SPL}.

\subsection{Robust FCs generation}\label{Sec:Universality}
The excited and propagated NSPPs are described by $u(x,t)=|u(x,t)|\exp\{\text{i}\phi_\text{NL}-\bar{\alpha}(\omega)x\}$; $\bar{\alpha}(\omega)=\text{Im}(\mathcal{K}(\omega))$, possess self-focusing nonlinearity and weak second-order dispersion that affects the plasmonic pulse envelope. The NSPPs thereby excite as plasmonic peregrine wave and Akhmediev breather and  their phases undergo nonlinear dynamical evolution that propagates as modified pattern, as it is represented in Figs.~\ref{Fig:Two}(a) and (c). Due to SPP pulse compression, a maximum plasmonic field intensity $|u(x,t)|^2\mapsto u_\text{max}$ and two dark points $|u(x,t)|^2\mapsto 0$ is expected due to growth-return cycle for both NSPP waves. The SPP phase corresponds to these dark points are singular, as we depict in Figs.~\ref{Fig:Two}(a) and (c), which we term as phase singularities~(PSs). To achieve PSs, we assume input plasmonic field as an evanescent wave with input power $P_\text{p}$ characterized by $\mathcal{U}_{0}(x=0,t)=\sqrt{P_\text{p}}\exp\{\text{i}\theta_{l}-\bar{\alpha}x\}$ and consider the seeded noise as a perturbation with amplitude $u_\text{N}=0.08u_{0}$ and modulation frequency $\nu_\text{mod}$ as $\Delta u_\text{N}=u_\text{N}\cos[2\pi\nu_\text{mod}t]$ that introduces a small perturbation to plasmonic field. We then achieve the SPP dynamics by numerically solving the nonlinear Schr\"odinger equation~\cite{PhysRevA.98.013825,PhysRevA.99.051802,Asgarnezhad-Zorgabad:20} for $u(x=0,t)=\mathcal{U}_{0}(x=0,t)+\Delta u_\text{N}$. Next, we investigate the amplitude of the plasmonic peregrine wave and Akhmediev breather in Fourier space in Figs.~\ref{Fig:Two}(b) and (d).

Various nonlinear plasmonic phases such as periodic~(Fig.~\ref{Fig:Two}(a)) and single PSs~(Fig.~\ref{Fig:Two}(c)) are excited by tuning the modulation parameter through plasmonic Akhmediev breather and peregrine wave formation, respectively. Figs.~\ref{Fig:Two}(b) and \ref{Fig:Two}(d) demonstrate that the FCs correspond to these PSs generate for both NSPP excitation and these quantities are invariant against input field modulation. Consequently, PSs are the \textit{similar} features of the exciting nonlinear waves, for peregrine wave and Akhmediev breather. For a characteristic frequency $\omega_\text{ch}=10~\text{MHz}$, robust FCs up to $\omega_\text{comb}\approx 3\omega_\text{ch}$ are achieved through plasmonic PS. Consequently, similar frequency comb generation and their stable propagation through interaction interface are referred to as invariant of NSPPs. The FCs $|\omega|<2\omega_\text{ch}$ can propagate for a few propagation lengths $-5.5L_\text{NL}<x<-4L_\text{NL}$ and hence would produce a robust plateau, as it is shown clearly in Figs.~\ref{Fig:Two}(b) and \ref{Fig:Two}(d), which we exploit this square to design a plasmonic version of SL. 

\begin{figure}[t]
\centering
\fbox{\includegraphics[width = 1\linewidth]{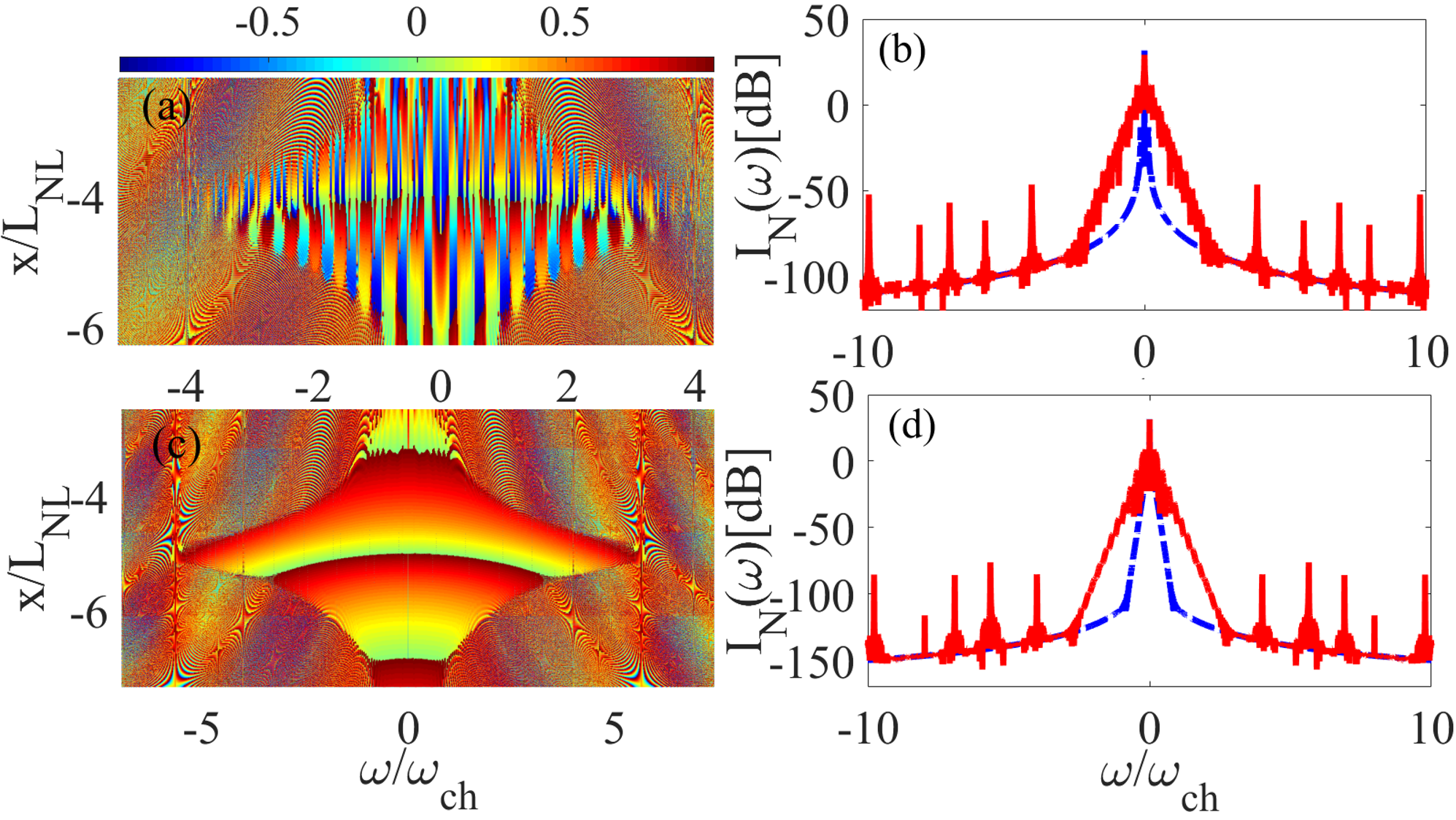}}
\caption{Spectral evolution of NSPPs through the interaction interface: Panel (a) is the spectral phase variation $\phi(x,\omega)$ panel (b) represents the logarithmic spectral harmonic intensity of the plasmonic breather as a function of perturbation frequency. Panel~(c) denotes the phase variation and (d) is spectral logarithmic power density for plasmonic peregrine wave excitation. In both panels (b) and (d) the blue dotted-dashed line represents the input field and red solid-line denotes the NSPP excitation in the presence of invariants~[Eq.~\eqref{Eq:Invariant_Nonlinear}]. Despite the dissimilar phase variation, excited FCs are the invariant of the nonlinear system. See the text for more details.}
\label{Fig:Four}
\end{figure}

\subsection{Invariants of nonlinear system}\label{Sec:Invariants}
The formation of robust PFCs and their robustness against external field modulation can be elucidated through the apparition of hidden invariants of this nonlinear system, which we termed invariant parameters. One of the key theoretical results of this work is that the PFCs act as invariant features across different types of nonlinear waves excitation. PFCs are invariant of this plasmonic system if we fulfill two conditions simultaneously, namely, (i) $\mathcal{N}$, which is the number of plasmonic frequency combs, remains as constant along interaction direction ($x$) for any kind of nonlinear plasmonic waves, and, (ii) the output spectral envelope $I_\text{N}(\omega)$ after a few nonlinear propagation lengths would be the same for both plasmonic peregrine and Akhmediev breather. In this section we establish that the PFCs stably propagate in the interaction interface and obtain a modified nonlinear evolution equation, that can be used for different classes of NSPPs in the presence of conservative conditions. In \S~\ref{Sec:SPL}, we prove the similarity between the spectral evolution of the plasmonic peregrine wave and Akhmediev breather and justify the formation of invariant PFCs.

We note that our robust FCs propagate for characteristic length $L_\text{eff}=\bar{\alpha}^{-1}[1-\exp\{-\bar{\alpha}L\}]$ and within nonlinear timescale $t_\text{S}=\tau_0+\partial_{\omega}[\ln\{(n_\text{eff}S_\text{eff})^{-1}\}]$. To achieve the invariant parameters, we evaluate the spatial variation of the energy flux and number of plasmon modes associated with NSPPs within spectral domain. In our analysis, we consider the FCs correspond to central SPP modes $\omega_\text{SPP}$, $\omega_\pm=\omega_\text{SPP}\pm\omega_{m}$ situated within the electromagnetically induced transparency window commensurate with the suppressed Ohmic loss. The nonlinear coefficient related to transparency window is $\mathcal{W}(\omega_l)$, we introduce $\Delta:=4\text{Im}[A_{1}^{*}A_2A_3A_{4}^{*}\exp\{\text{i}\Delta\mathcal{K}_\text{t}x\}]$ as SPP field detuning for four-wave mixing process, define phase-matching parameter $\Delta\mathcal{K}_\text{t}=\mathcal{K}(\omega_\text{ch})+\mathcal{K}(\omega_{+})-\mathcal{K}(\omega_{0})-\mathcal{K}_4(\omega_{-})$; $\mathcal{K}(\omega_{l})=\beta(\omega_{l})+k(\omega_l)+\sum'\sum_{n}\left(2-\delta_{ln}\right)\mathcal{W}(\omega_{l})|A_n(\omega_l)|^2$, consider effective nonlinear parameter as $\mathcal{W}_\text{eff}=\mathcal{W}(\omega_\text{ch})+\mathcal{W}(\omega_{0})-\mathcal{W}(\omega_{-})-\mathcal{W}(\omega_{+})$ and consider $k_l$ as corresponding wavenumber of the propagated modes. Following the technical details of derivation as in \S~S.3 A2, we evaluate the spatial dynamics of energy flux as
\begin{align}
	\frac{\partial E}{\partial x}\propto&-\mathcal{O}(\bar{\alpha})+\sum_{m}\mathcal{W}_\text{eff}\Delta,
\end{align}
and number of stable plasmon modes as
\begin{align}
	\frac{\partial\mathcal{N}}{\partial x}\propto&-\mathcal{O}(\bar{\alpha})+\sum_{m}\left[\frac{\mathcal{W}(\omega_{0})}{2\omega_0}+\frac{\mathcal{W}(\omega_{-})}{\omega_0+\omega_{-}}+\frac{\mathcal{W}(\omega_{+})}{\omega_0+\omega_{+}}\right]\Delta.
	\label{Eq:Spatial_Nonlinear_Modulation}
\end{align}

The dynamical evolution of the PFC, their robustness, and conservative conditions hence are limited by the nonlinear coefficient $\mathcal{W}(\omega)$ and loss of the system. We introduce loss suppression and careful nonlinear modulation are dual requirements for conservation of energy and number of plasmon modes and hence, we expect robust FCs generation only for ultra-low interface $\bar{\alpha}\mapsto0$, and for finely tuned nonlinear coefficient. To characterize modified nonlinearity, we consider the nonlinear coefficient as $\mathcal{W}(\tilde{\omega})=\mathcal{W}_{0}+\mathcal{W}_1\delta\tilde{\omega}+\mathcal{W}_2\delta\tilde{\omega}^2+\mathcal{O}(\delta\tilde{\omega}^3)$. The energy can be a conservative quantity of the system~(i.e. $(\partial E/\partial x)\approx0$) by taking $\mathcal{W}(\tilde{\omega})=\mathcal{W}_{0}+\mathcal{W}_1\delta\tilde{\omega}$ whereas we achieve the conservation of the excited FCs~(i.e. $(\partial\mathcal{N}/\partial x)\approx0$) for $\mathcal{W}(\tilde{\omega})=\mathcal{W}_{0}+\mathcal{W}_1\delta\tilde{\omega}+\mathcal{W}_2\delta\tilde{\omega}^2$. Consequently, the energy flux and number of plasmonic modes are simultaneous invariants of the system for
\begin{equation}
	\mathcal{W}(\tilde{\omega})=\mathcal{W}_0\left(1+\frac{\tilde{\omega}}{\omega_0}\right).
	\label{Eq:Invariant_Nonlinear}
\end{equation}
This equation establishes that the apparition of conservation within a plasmonic system are independent of the SPP field dispersion/dissipation. The conservation of PFCs across different classes of NSPPs is an essential step to justify the invariance of FCs within our system~(see \S~\textcolor{blue}{C 1b} of appendix for more details). In what follows, we describe the nonlinear spectral evolution of the NSPP field in the presence of conserved $\mathcal{N}$, $E$, and employ our modified nonlinear equation to establish the apparition of invariant PFCs.

\begin{figure}[t]
\centering
\fbox{\includegraphics[width = 1\linewidth]{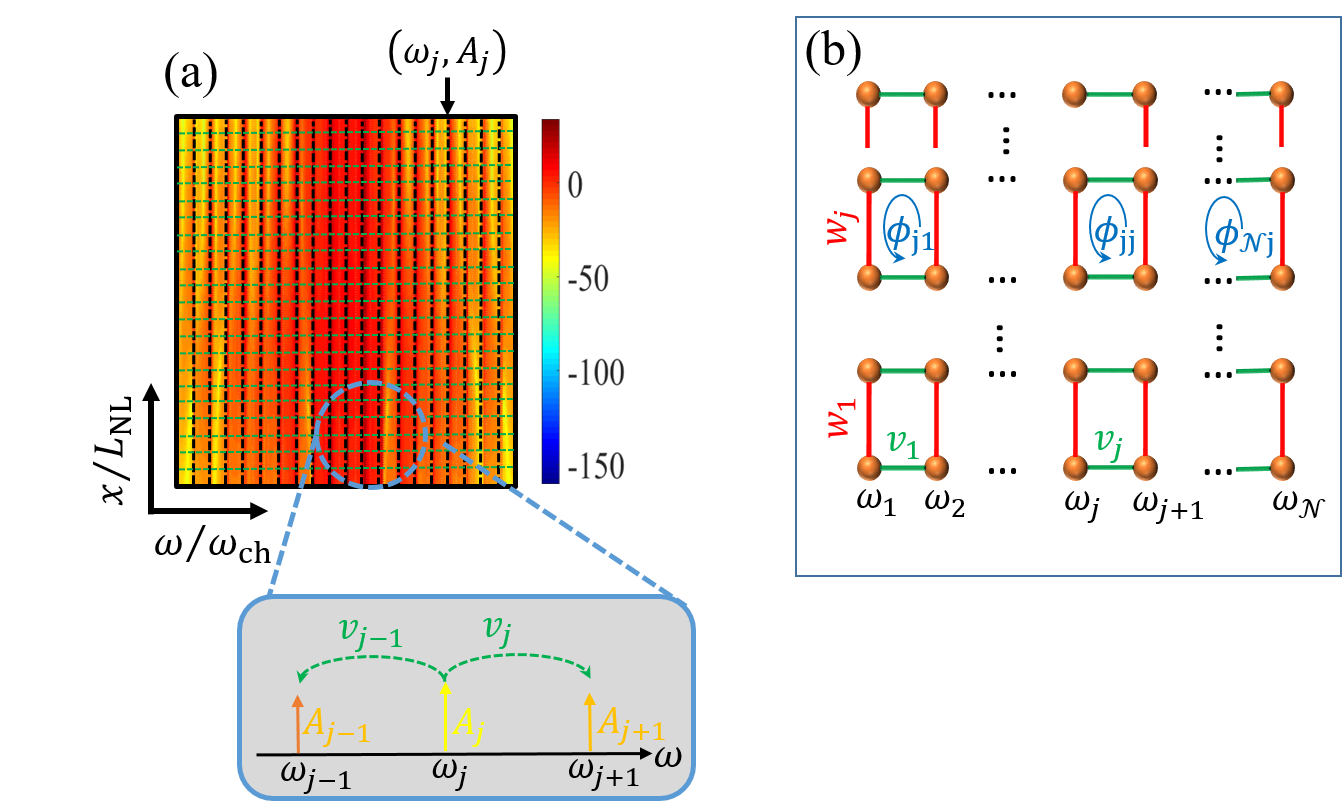}}
\caption{Mapping between the stable FCs to a synthetic dimension: Panel (a) represents the power spectrum of the spectral harmonic side-bands for $|\omega|<2\omega_\text{ch}$ and $-5.5L_\text{NL}<x<-4L_\text{NL}$ correspond to robust propagation of FCs. the inset of this figure represents the correlation between the FCs. Panel (b) is the qualitative description of the synthetic lattice corresponding to frequency comb excitation.}
\label{Fig:Three}
\end{figure}

To describe the NSPP evolution, we should solve nonlinear Schr\"odinger equation considering conservative conditions and by assuming nonlinearity as Eq.~\eqref{Eq:Invariant_Nonlinear}, which is challenging due to the plasmonic field being propagated within the dispersive interface. To remedy this limitation, we consider SPP field as electromagnetic waves that can be quantized through interaction interface and employ the quantum SPP approach that is independent of interface dispersion. Here we consider the NSPP field as time-harmonic profile $\bm{\mathcal{J}}(\bm{r},t):=\bm{\mathcal{J}}_{\omega}\exp\{\text{i}\omega t\}+\text{c.c.}$; $\bm{\mathcal{J}}\in\{\bm{E},\bm{B}\}$, next, we consider Weyl gauge field and evaluate vector potential $\bm{A}(\bm{r},t)$. Following Ref.~\cite{quantization}, we expand this potential in terms of amplitude and frequency of each mode and evaluate the time average classic energy for this SPP field. The quantum counterpart of this energy yields the quantization of PFCs with bosonic annihilation creation operators $\hat{\bm{b}}$~($\hat{\bm{b}}^{\dagger}$)~\footnote{As SPP wave conservatives are independent of interface dispersion, without the loss of generality, we consider $\mathcal{K}(\omega)=0$.}.  We consider the nonlinear coefficient of the interface as $\mathcal{W}^0$, and rewrite the interaction Hamiltonian for the stable nonlinear quantum plasmon mode as $\mathcal{H}_\text{I}=\sum_{m}\iint\;\text{d}\tilde{\omega}\text{d}\omega(\mathcal{W}^0/2)\hat{\bm{b}}_{\omega}^{\dagger}\hat{\bm{b}}_{\tilde{\omega}}^{\dagger}\hat{\bm{b}}_{\tilde{\omega}-\omega_m}\hat{\bm{b}}_{\tilde{\omega}+\omega_m}$. Next, we use Eq.~\eqref{Eq:Invariant_Nonlinear} and employ the Heisenberg equation of motion~\cite{scully1999quantum} to evaluate the dynamics of the mean-field value associated with the stable plasmon mode~($\partial\braket{\hat{A}_m}/\partial x$)~\cite{PhysRevA.40.844}, and include the dispersion term due to plasmonic field. We then achieve the evolution of NSPPs similar to Eq.~\eqref{Eq:Dynamics_SPP_Field} but with including invariant parameters as
\begin{equation}
	\frac{\partial\tilde{A}}{\partial x}=\text{i}\mathcal{K}(\omega)\tilde{A}+\sum_m\mathcal{F}\left[\Lambda(\tilde{\omega})|A_m|^{2}A_m\right]+\text{c.c.},
	\label{Eq:Modified_NLSE}
\end{equation}
for $\mathcal{F}$ the Fourier transform operator and $\Lambda$ is the nonlinearity that is linearized in the presence of conserved energy and invariant number of excited SPP modes~(see supplementary material \S~\textcolor{blue}{C~2}). 

\emph{Discussion on the universality of FCs}- In this work, we uncover the invariants only for plasmonic peregrine wave and Akhemediev breather, which means that our approach falls short of providing a universal description of nonlinear SPPs and frequency comb generation. Here, PFCs propagation and SPL formation are highly limited by loss suppression and nonlinear modification\footnote{We termed this dual requirement as conservative condition}. Consequently, our predicted effects would depend on robust PFC generation and conservation conditions, and hence are device-dependent. The extension of stable FC formation to other classes of solitons needs careful investigations and goes beyond the scope of this work.

\subsection{Nonlinear spectral dynamics and synthetic lattice formation}\label{Sec:SPL}
In this section, first, we achieve the output spectral envelope function of the NSPP waves for peregrine and Akhmediev breather cases with and without the existence of conservatives, to establish the invariance of PFCs in our hybrid system that can be interpreted as a synthetic dimension. Next, we exploit this synthetic space to construct an SPL.

To investigate the spectral dynamics and evaluate the output envelope profiles of the NSPPs, we assume SPP field as $A(x,\omega):=\sqrt{P}_\text{p}\exp\{\text{i}\phi_\text{S}-\text{i}\mathcal{K}(\omega)x\}$ and numerically solve Eq.~\eqref{Eq:Modified_NLSE} considering $x_0=-10L_\text{NL}$. For a lossy plasmonic interface with $\chi^{(3)}(\omega):=\mathcal{W}^0$, the plasmonic peregrine waves and Akhmediev breather have different output spectral envelope, as it is clearly shown in Figs.~\ref{Fig:Four}(b), (d)~(dashed blue curves), indicating that in the interaction interface with constant nonlinearity, PFCs are not robust and invariant of the system. We also investigate the spectral-spatial dynamics of phase $\phi(x,\omega)$~(Fig.~\ref{Fig:Four}(a) for Akhmediev breather and Fig.~\ref{Fig:Four}(c) for plasmonic peregrine wave) and output spectral envelope of the corresponding NSPP waves in the presence of conservation conditions, representing that the output spectral profile of the NSPPs for both plasmonic Akhmediev breather (Fig. \ref{Fig:Four}(b) red curve) and peregrine wave~(Fig. \ref{Fig:Four}(b) red curve) becomes the same.     

\begin{figure}[t]
\centering
\fbox{\includegraphics[width = 1\linewidth]{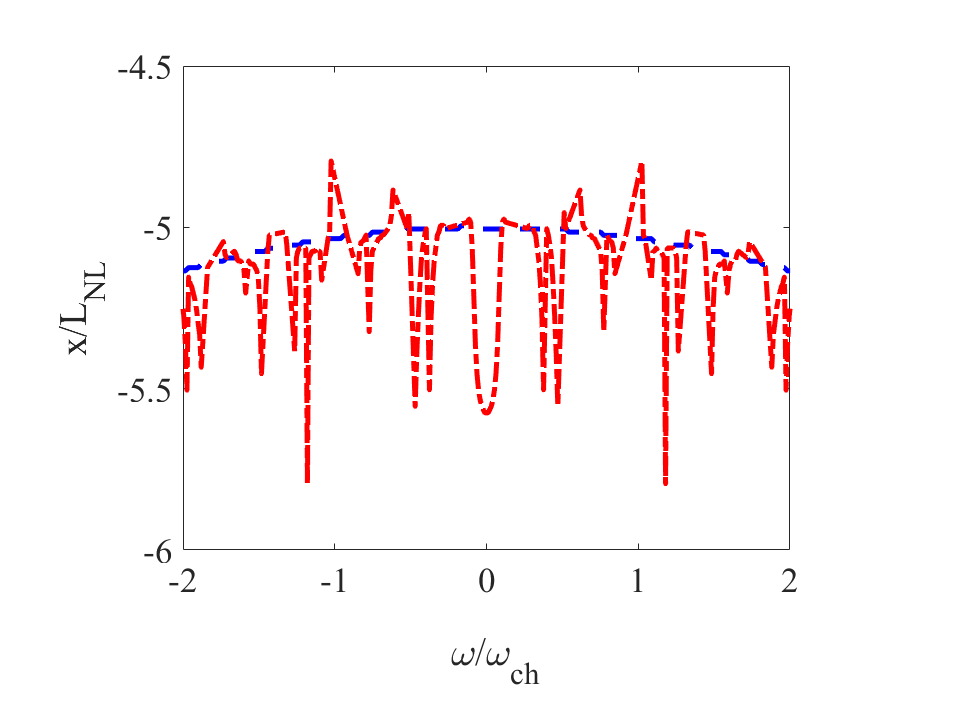}}
\caption{Observation of anomalous phase hopping, and non-zero phase trajectories through different NSPP field excitation: Dashed blue curve is the trajectory of non-zero phase hoppings $\phi_{i,j}\neq0$ for the plasmonic rogue wave formation, and red dashed-dotted curve is the corresponding phase trajectory for the breather formation.}
\label{Fig:Five}
\end{figure} 

Our results hence indicate that in a nonlinear waveguide fulfilling the vanishing loss and tunable nonlinearity, $(\mathcal{N},E)$ are conserved parameters of the systems. In this case, the interaction interface supports the stable propagation of plasmonic peregrine wave and Akhmediev breather for a few propagation lengths with similar output spectral probe envelope. Consequently, FC is the hidden conserved parameter among both NSPPs, and propagated PFCs are interpreted as a synthetic dimension in our scheme. We exploit this synthetic dimension to construct an SPL. We present our qualitative and quantitative approach towards SPL formation in two steps, namely, (i) well-defined synthetic dimension, (ii) constructing synthetic lattice. In what follows we describe these two steps in detail.
	
\emph{Qualitative description}- Qualitatively, we prove that PFCs are invariant of excited NSPPs and they propagate as discrete frequency components with equal frequency spacing $\delta:=\omega_\text{EIT}/\mathcal{N}$, they act as an internal degree of freedom for SPPs and can be considered as a synthetic dimension~\cite{FanReview2018}. Consequently, we assume $x/L_\text{NL}$ and $\omega/\omega_\text{ch}$ to construct SPL as it is qualitatively represented in Fig.~\ref{Fig:Three}(b). To SPL formation, we need to characterize lattice sites and hopping between these sites. In our scheme, discrete FCs effectively act as lattice sites, and we consider SPP field overlapping between FCs and their nearest neighbor as hopping between PFCs, which qualitatively describes SPL formation. As these combs are stable for a limited propagation length, our synthetic structure can be considered as a two-dimensional lattice in $x$-$\omega$ plane. 

\emph{Quantitative description}- We assume PFCs excite with frequency components $\omega_n=\pm n\delta$, and with propagation constant $\xi_n=\xi_\text{NL}+\Delta\xi_\text{L}$; $\xi_\text{NL}=\sum_n\mathcal{C}_n\cos(k_n)$ is the nonlinear part of propagation constant for $\mathcal{C}_n:=\mathcal{W}\sum_m A_m(0)A_{m-n}^{*}(0)$ the correlation between different harmonic side-bands. We assume $\mathcal{W}$ as Eq.~\eqref{Eq:Invariant_Nonlinear} to include the conservative conditions and consider only the nearest neighbor effect in our analysis. The total electric field for each PFCs are then $|\bm{E}|=\sum a_n(x)\exp\{\text{i}[\omega_n t-\xi_n x]\}$. We then achieve the phase matching condition for PFCs in terms of system parameters as $\Delta\xi_\text{L}=\delta n_\text{g}/\text{c}$\footnote{c is the speed of light in vacuum.}. These characterized PFCs then act as a synthetic dimension for two NSPP waves. The dynamical evolution of $a_n$ is then achieved through coupled-mode equation $\partial_x a_n(x)=-(\alpha_\text{eff}/2)a_n(x)+\text{i}[\mathcal{C}a_{n+1}(x)-\mathcal{C}^{*}a_{n-1}(x)]$. Following Ref.~\cite{FanReview2018}, the Hamiltonian describing this coupled-mode equation is $\mathcal{H}_0=\kappa\sum_n[\ket{n+1}\bra{n}+\ket{n-1}\bra{n}]$. (see further mathematical details in\S~S.4). To extend this Hamiltonian to a two-dimensional case, we need to consider hopping along x-axis as well. Within a characteristic $x$-$\omega$ plane presented in Fig.~\ref{Fig:Three}(a), we assume that the amplitude of PFCs are robust and consequently, we consider the phase variation along this direction.  

We write the Hamiltonian of SPL in $x$-$\omega$ plane in terms of hoppings $w_{i,j}$ and $v_{i,j}$ in the most general case\footnote{$i\in\{-\mathcal{N},\ldots,\mathcal{N}\}$ and $j$ are dummy variables along $\omega$ and $x$ directions respectively, and $\{n\}$ represents any set of nighbors along the frequency direction, however we only consider $n=1$.}
\begin{equation}
    \mathcal{H}_\text{SPL}=\sum_{i,j,n}w_{i,j}\hat{a}_{i,j}\hat{a}_{i+n,j}^{\dagger}+\sum_{i,j}v_{i,j}\hat{a}_{i,j}\hat{a}_{i,j+1}^{\dagger}+\text{H.C.},
    \label{Eq:Hamiltonian_Square}
\end{equation}
and consider the evolution of this lattice along the interaction interface as $\ket{\psi(x)}=\exp\{\text{i}\mathcal{H}_\text{SPL}x\}\ket{\psi(x=x_0)}$. This system is a synthetic lattice corresponding to the nonlinear interaction within a plasmonic interface, hence its dynamics are described through the annihilation-creation operators $\hat{a}$~($\hat{a}^{\dagger}$), and complex hoppings $w_{i,j}$, $v_{i,j}$. To achieve the SPL Hamiltonian, we also include the dissipation to the SPP wave dispersion $\mathcal{K}(\omega)\mapsto\mathcal{K}(\omega)+\text{i}\bar{\alpha}$. Eq.~\eqref{Eq:Hamiltonian_Square} indicate that SPL has square-type structure but with complex coefficients that yield \textit{anomalous} hopping phase between lattice sites $i,j$, we termed as $\phi_{i,j}$, which is non-zero for specific plateau within our SPL. For our nonlinear plasmonic system, this hopping phase depends on the modulation parameter, is different for various nonlinear field excitation and we achieve this non-zero flux~($\delta\phi_{i,j}\neq0$) only for specific spatial-spectral trajectories, as it is clearly shown in Fig.~\ref{Fig:Five}. These trajectories are reciprocal, frequency dependent and can be exploited to introduce anomalous artificial gauge field, and is suitable to produce non-uniform synthetic magnetic fields.

\section{Conclusion}\label{Sec:Conclusion}
To sum up, we exploit the spectral SPP field evolution to discover the hidden invariant parameter of the nonlinear plasmonic wave systems. The key theoretical result of this work is that in a plasmonic waveguide with suppressed loss and tunable nonlinear coefficient, PFCs act as an invariant across different types of nonlinear waves and are interpreted in terms of a synthetic dimension to construct an SPL. Invariant PFCs apparition and SPL formation are based on assumptions related to nonlinear waveguides and assumptions related to the SPP field. For the SPP field, we assume far-field approximation, neglect its phase variation, and consider the effect of evanescent coupling through mean-field averaging. For plasmonic configuration, we assume the interface with low dispersion and controllable nonlinearity. Our loss compensation scheme and field propagation methodology in our proposed scheme is then an extension to the ultra-low loss plasmonic scheme represented in~\cite{PhysRevA.99.051802,Asgarnezhad_Zorgabad_2020,Asgarnezhad-Zorgabad:20} that the gain compensated graphene structure play the role of the metamaterial layer. In particular, we uncover that the FCs and PSs are similar features for plasmonic peregrine waves and Akhmediev breather.

Next, we prove that the linearized nonlinear coefficient provides similar PFCs that can be interpreted as the internal degrees of freedom and thereby act as synthetic dimensions. Qualitatively, we achieve this finely tuned nonlinearity as control parameters. Our analysis thereby indicates that dual requirement of the vanishing losses, linearizing the interface nonlinearity, and the existence of robust FCs with preserved number of excitation are requirements to constructing an SPL. We quantitatively describe the robust FCs generation through quantum NSPP field formalism commensurate with the Schr\"odinger approach and we also perform the mean-value averaging to achieve the Fourier dynamics of NSPPs in the presence of conservative conditions. Our approach justifies the existence of the anomalous hopping phase through characteristic reciprocal trajectories, which depend on input field modulation and can be exploited to various effects, from nonuniform magnetic flux to anomalous gauge field.

\pagebreak
\newpage
\setcounter{equation}{0}
\setcounter{figure}{0}
\setcounter{section}{0}
\setcounter{page}{1}
\renewcommand{\theequation}{S\arabic{equation}}
\renewcommand{\thefigure}{S\arabic{figure}}
\renewcommand{\thesection}{S.\arabic{section}}
\newcounter{SIfig}
\renewcommand{\theSIfig}{S\arabic{SIfig}}
\begin{widetext}
\section*{Supplementary information: Synthetic plasmonic lattice formation through invariant frequency combs excitation in coherent graphene structures}
\begin{quote}
\centering
    \normalsize Zahra Jalali-Mola$^1$ and Saeid Asgarnezhad-Zorgabad$^{1,*}$\\
    $^1$\small\textit{Independent Researcher, Personal Research Laboratory, West Ferdows Blvd., Tehran 1483633987, Iran}\\
    \footnotesize $^{*}$\textcolor{blue}{sasgarnezhad93@gmail.com}
\end{quote}
\begin{quote}
    We present the quantitative and qualitative details towards synthetic plasmonic lattice formation in this supplementary information. We start by presenting an example of the nonlinear waveguide, comprising a coherent multi-level atomic medium situated on top of an ultra-low loss double graphene layer. Specifically, we elucidate the specific mechanisms required to excite and probe the system within this design, the multi-level structure, and transitions of the $^{87}$Rb atoms. We also represent the detailed explanation of scheme feasibility by introducing a realistic source-waveguide-detection triplet in \S~\ref{Sec:Realistic_Waveguide}. Next, we present the detailed quantitative description of our double graphene layer, its optical properties and Ohmic-loss compensation through gain-loss modulation in \S~\ref{Sec:Linear_Response}. Then, in \S~\ref{SM:Invariant_evolution} we provide a detailed mathematical steps towards spectral dynamics of the SPP pulse through the hybrid interface and finally in \S~\ref{Sec:Construction_synthetic} we elucidate the additional quantitative steps towards the synthetic plasmonic lattice formation.
\end{quote}
\section{\label{Sec:Realistic_Waveguide}Detailed description of the waveguide configuration: An example with multi-level atomic medium}
In this section, we proceed with our approach, by elucidating a specific example of the nonlinear waveguide that meets our conservation conditions, which comprises an atomic ensemble situated on top of the ultra-low loss double-layer graphene. Consequently, first, we present all detailed explanations of the waveguide operation in \S~\ref{Sec:Realistic_Waveguide_Detailed}, and next in \S~\ref{Sec:Realistic_Waveguide_Realistic}, we explain the experimental feasibility for this specific waveguide design. Finally, we discuss the system parameters and challenges of our suggested setup in \S~\ref{Sec:Realistic_Waveguide_Discussion}.

\subsection{\label{Sec:Realistic_Waveguide_Detailed}Detailed explanation of waveguide operation}
To excite invariant PFCs and establish SPL formation, the nonlinear waveguide should possesses suppressed loss and tunable nonlinearity. The dual conditions of loss compensation and tunable nonlinear control can be achieved in an atomic ensemble irradiated with laser fields. The proper injection of laser fields would yield formation of electromagnetically induced transparency windows, for which, the electronic transitions of atomic medium interfere to eliminate or reduce resonant atomic absorption and modulate the linear dispersion. Atomic medium with this spectral transparency windows possesses ultra-low linear absorption and giant nonlinear coefficient. The width related to transparency window nonlinear parameters off this atomic ensemble can be controlled via intensity and detuning of the driven laser fields. Consequently, nonlinear waveguide with double graphene layer as bottom layer and multi-level atomic medium situated on top can be a possible candidate to achieve invariant PFCs and SPL formation. In what follows, we describe the details of the waveguide configuration in details.

To uncover the invariant parameters of NSPPs, we suggest a plasmonic nanostructure as shown in Fig.~\ref{fig:Fig_One}. This waveguide comprises three parts (i) source, (ii) waveguide, and (iii) detection. On one end of the waveguide, a fiber-based connector is attached to couple the source fields to the waveguide, and on the other end, a detection system is connected to detect the output SPP waves. The source fields produce SPPs, plasmonic waveguide controls their spatiotemporal profile that yields linear/nonlinear SPP generation and the detector collects the output intensity of plasmonic fields.

\begin{figure}[t]
\centering
\fbox{\includegraphics[width=1\linewidth]{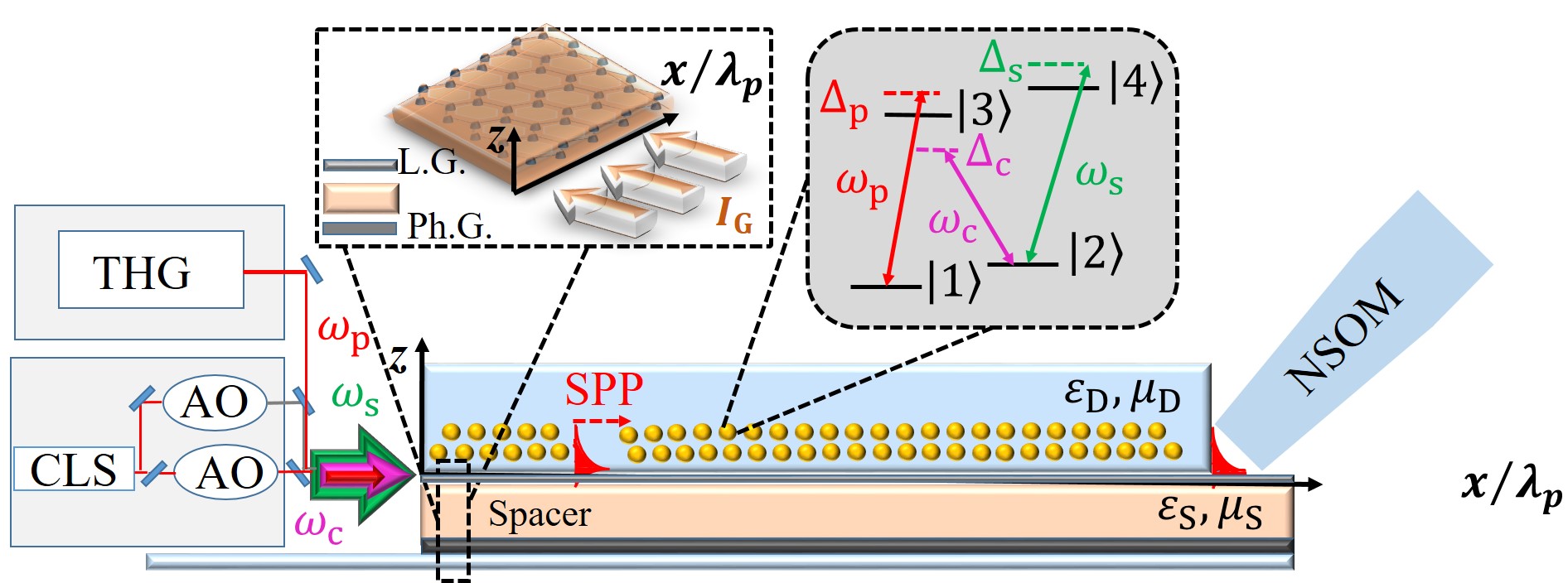}}
\caption{ Source-waveguide-detection triplet to construct an SPL: CLS, and THG irradiate the system and act as \textit{source}. The waveguide comprises cold atomic gas situated on top a double-layer graphene structure. Detection system is attached as an atomic force microscope tip at the end of the waveguide. Right inset represents the atomic medium transitions and the left inset is the double-layer graphene structure design. Detuning frequencies are $\Delta_{l}$ and Rabi frequencies are $\Omega_{l} \propto\bm{E}_{l}$; $l\in\{\text{c},\text{s},\text{p}\}$. The inset also shows the atomic states $\ket{j}$ and the details of the coupling mechanism. The symbols are: THG:terahertz generator, CLS: coherent laser source, AO: acousto-optic modulator, L.G: lossy graphene, Ph.G: photo-inverted graphene.} 
    \refstepcounter{SIfig}\label{fig:Fig_SM_One}
\end{figure}

\emph{General description}- Coherent laser source~(CLS) and terahertz generator~(THG) are attached to our waveguide using the end-fire coupling technique and irradiate hybrid structure as field sources. THG produces a weak SPP field that is resonant with $\ket{1}\leftrightarrow\ket{3}$ dipole transition~\cite{Sfei2012gate} and CLS provides lights that coupled atomic transition as represented in Fig.~\ref{fig:Fig_SM_One}. We employ controllability over intensities and detunings of fields to suppress losses and controlling nonlinearity through the interaction interface. Finally, we detect NSPPs at the output using NSOM technique. In our analysis, we consider a strong couple~(c), a signal~(s), and a terahertz probe~(p) as driving fields that are linearly polarized, possess temporal coherence longer than the waveguide decay, and longitudinally coherent enough to cover the waveguide. The couple and signal laser are obtained through a dye laser and we also control their frequencies through acousto-optic modulators.

On the other hand, our hybrid waveguide comprises a graphene multi-layer as a bottom medium and a cold atomic gas as the upper layer~(see Fig.~\ref{fig:Fig_SM_One}). We consider graphene-dielectric-graphene as a multilayer structure that possesses ultra-low loss for our SPP wavelength of interest. Above this structure, an ensemble of $^{87}\text{Rb}$ cold atoms are situated. Specifically, we consider $D$ line of $^{87}\text{Rb}$ atoms as four-level N-type atomic medium~(4NA) with
\begin{align}
    \ket{1}=\ket{5^{2}S_{1/2},F=1},\qquad \ket{2}=\ket{5^{2}S_{1/2},F=2},\nonumber\\
    \ket{3}=\ket{5^{2}P_{3/2},F=2},\qquad
    \ket{4}=\ket{5^{2}P_{3/2},F=3},
\end{align}
as transition levels. Atomic density of this ensemble is $N_\text{a}$, homogeneous decay rate of the $\ket{m}\leftrightarrow\ket{n}$ transition is $\gamma_{nm}$, dephasing rate is $\gamma_{nm}^\text{deph}$, and the corresponding atomic dipole moment is $\bm{d}_{nm}$.

\emph{Design of the waveguide\label{Waveguide design}}- To sum up, our structure is a specific design that includes cold 4NAs and multi-layer graphene structure. 4NAs are appealing due to suppressed absorption commensurate with giant Kerr nonlinearity, thereby this medium provides stable NSPP propagation and FCs generation. On the other hand, we exploit multilayer graphene due to its stable optical properties against the external magnetic field gradient, giant field concentration, and opportunities to provide ultra-low loss plasmonic field propagation. Consequently, our process is coherent, possesses ultra-low loss, and is suitable to provide control over optical nonlinearity. These conditions are necessary to provide suppressed loss and careful nonlinear modification, which are dual requirements for robust PFCs propagation and conservative conditions, and hence our design are the best candidate for SPL formation.

\subsection{\label{Sec:Realistic_Waveguide_Realistic}Realistic description of the waveguide}
In this section, we provide a detailed discussion of the experimental feasibility of our waveguide. As we elucidate in \S~\ref{Sec:Realistic_Waveguide_Detailed}, our waveguide comprises of three parts, namely, (i) source, (ii) waveguide, and (iii) detection. In what follows, we explain the feasible experimental implementation of these parts.

\emph{Source}- As for sources, two laser fields, a strong couple~(c), a signal~(s), a terahertz weak probe~(p) field, and an additional trigger laser, drive the waveguide. Signal, couple, and probe fields the same polarization and are obtained from an external cavity diode laser that is narrow-band, frequency stabilized, linearly polarized, temporally longer than the waveguide decay, and longitudinally coherent enough to cover the waveguide~\cite{Das2019s}. The frequency of the source fields is modified using acousto-optic modulators. Moreover, we propose generating probe field with linear polarization in nanoscopic scale by assuming oxide nanojunctions that is suitable for producing terahertz radiation using ultrafast frequency mixing~\cite{Sdoi:10.1021/nl401219v}. This probe wave will produce the SPP field with a wavenumber $\omega_\text{p}$ that is in resonance with atomic dipole transition wavelength~\cite{Sfei2012gate}. Forth field is a pulsed femtosecond fiber-laser with a few $\text{MHz}$ repetition rate, which is linearly polarized, acts as a trigger field, irradiates the plasmonic waveguide from the bottom side and injected perpendicular to the driving laser fields to produce gain for graphene layer~\cite{SWatanabe_2013}. 
the couple and signal fields are injected to the waveguide and co-propagate parallel to the emitter-graphene interface using end-fire coupling technique~\cite{SPhysRevB.63.125417}.

\emph{Waveguide}- Our waveguide comprises a thin layer atomic medium doped on a lossless dielectric situated on top of a hybrid nanostructure. This plasmonic apparatus comprises two parts. A thin layer foil as a bottom medium that serves as a holder and a double-layer graphene scheme, which is a graphene-spacer-graphene multilayer. This plasmonic scheme should possess low-loss for the dipole transition wavelength. Various methods such as optimization/design, including virtual gain and parametric amplification serve to combat the loss of this plasmonic structure~\cite{SGhoshroy:20} and hence various ultra-low loss layers can be implemented as our dispersive layer. However, this layer should be robust against magnetic gradient that should be employed for atom cooling.

Graphene structures can be a potential candidate as a metallic-like layer, due to loss tunability and wide spectral bandwidth but producing resonant excitation of NSPPs in optical graphene within a single layer of this structure, unfortunately is challenging due to Ohmic loss and the need for high doping level~\cite{Srodrigo2017double}. To remedy these limitations, we suggest a double- or multi-layer graphene structures, which is experimentally verified and equivalently act as a single-layer graphene with suppressed dissipation and with high-doping. Our graphene reconfiguration is robust to low magnetic gradients and hence is a suitable candidate for this scheme. We introduce trigger laser to the bottom graphene layer to induce photo-inverted gain~\cite{SHess2015_1} that is exploited to suppresses the Ohmic loss related to this plasmonic structure. Consequently, this configuration would be ultra low-loss for the dipole transition wavelength and we expect stable propagation of linear/nonlinear SPPs within the atomic medium-plasmonic scheme interface.

The interaction interface is filled with a four-level $N$-type atomic gas~(4NA), that is cooled to ultra-low temperatures~\cite{SPhysRevA.58.3891}. This cold gas serves as electrical dipoles and we also assume the dopant thickness is a few dipole transition wavelengths. 4NAs are appealing due to its efficiency for providing controllable nonlinearity/dispersion and specifically, we consider $D$ line of $^{87}\text{Rb}$ atoms with $\ket{1}=\ket{5^{2}S_{1/2},F=1}$, $\ket{2}=\ket{5^{2}S_{1/2},F=2}$, $\ket{3}=\ket{5^{2}P_{3/2},F=2}$ and $\ket{4}=\ket{5^{2}P_{3/2},F=3}$ as transition levels. Atomic density is $N_\text{a}$, homogeneous decay rate of the $\ket{n}\leftrightarrow\ket{m}$ transition is $\gamma_{nm}$, dephasing rates are $\gamma_{nm}^{\text{deph}}$ and we neglect the inhomogeneous broadening due to weak Doppler effect. Our laser fields with Rabi frequencies $\Omega_{l}$ and correspond detuning frequencies $\Delta_{l}$; $l\in\{\text{c},\text{s},\text{p}\}$ drive the atomic medium through dipole approximation. We also assume these fields are tightly confined to the interactive interface with evanescent coupling function $\zeta_{l}(z)$~\cite{SPhysRevA.98.013825}.

\emph{Detection}- The nonlinear plasmonic processes and excited frequency combs are characterized using a detection system. To this aim, a sharpened multimode fiber is attached to the end of the plasmonic waveguide that is called \emph{tip}, and the plasmonic interface commensurate with tip is illuminated using an infrared focused beam. This field then interacts with NSPPs within the interaction interface, the scattered intensity field profile corresponds to this near-field is then propagates through the tip, the intensity pattern would collects using an image intensifier~\cite{SMorton:64} and would detect exploiting an atomic force microscope~\cite{Sfei2012gate}.

Our waveguide reconfiguration is consequently experimentally feasible from source to detection and is efficient to generate controllable linear and Nonlinear SPPs. We achieve the invariant parameters of the NSPPs in three steps. First, we obtain the spatiotemporal and spectral-spatial dynamics of the NSPPs in the presence of nonlinear parameters of the hybrid system. Next, we exploit the quantum properties of nonlinear gain modulation through the hybrid interface and establish a modified nonlinear evolution equation based on invariant parameters of this plasmonic scheme. Finally, we establish the robustness of frequency combs, number of excited plasmon modes, and phase singularity within the interaction interface. Using the robustness of frequency combs, we exploit a  synthetic lattice to elucidate the invariant of spectral dynamics and establish the apparition of the anomalous artificial gauge field.

\subsection{\label{Sec:Realistic_Waveguide_Discussion}Discussion on scheme feasibility}
Now, we present the system parameters for which the PFCs serve as invariant of the plasmonic system and can be interpreted as SPL. Then we discuss the challenges and outlook of our suggested scheme.

\emph{Scheme feasibility and simulation parameters}- First, we test the feasibility of our scheme. We assume $g_\text{s}=g_\text{v}=2$, electron Fermi velocity as $v_\text{F}=10^6~\text{m}/\text{s}$, chemical potentials for electron (hole) as $\mu_\text{e}=\mu_\text{h}=\mu/2\approx0.34~\text{eV}$, dielectric constant of spacer as $\varepsilon_\text{d}=2$, and $d\approx5~\text{nm}$. 4NAs are cooled to $T\approx 10~\text{mK}$ using a magneto-optic trap with $\text{d}B/\text{d}z=10~\text{G}/\text{cm}$ and we set $\lambda_\text{p}=1.55~\text{eV}$, $N_\text{a}=9\times10^{10}~\text{cm}^{-3}$, $\Omega_\text{c}\approx30~\text{MHz}$, $\Delta_\text{c}=-2~\text{MHz}$, $\Omega_\text{s}=35\text{MHz}$, $\Delta_\text{s}=16~\text{MHz}$ and $\Delta_\text{p}=0$ and choose relaxation for $^{87}\text{Rb}$ from Ref.~\cite{Ssteck2001rubidium}. The typical gradient magnetic field to change the optical response of graphene is a few $\text{T}$ for a $\text{cm}$ interaction length~\cite{SZhang2005} which is far from gradient field that is needed for magneto-optic trap and hence its effect on NSPP propagation is negligible. With these parameters a SPP with $\mathcal{K}_{2}=(-4.42+0.4\text{i})\times10^{-12}\text{s}^{2}\cdot\text{cm}^{-1}$, $W=(2.98+0.6\text{i})\times10^{-11}\text{s}^{2}\cdot\text{cm}^{-1}$, with group velocity $v_\text{g}=2\times10^{4}\text{m}/\text{s}$ propagates, we have $g_{\mathcal{K}_{2}}=1.01$, $f=0.045$, and $\text{Im}[K_\text{a}(\omega)]\approx0.06~\text{cm}^{-1}$. For gain graphene $\text{Im}[k_\text{G}]=-0.07~\text{cm}^{-1}$, and for our system $\text{Im}[k_\text{C}]=-0.02~\text{cm}^{-1}$, therefore $\bar{\alpha}=\text{Im}[K_\text{a}(\omega)+k_\text{C}(\omega)]=0.04\approx0$ that justifies ultra-low loss NSPP propagation.

Our waveguide consequently is suitable for ultra-low loss propagation of NSPPs within the atomic dipole transition wavelength. To simulate SPP propagation in this nonlinear plasmonic system, we assume (i) SPP waves propagate as far plasmonic fields, (ii) the plasmonic phases are constant~(i.e. $\mathcal{K}x-\omega t=\text{Const.}$), and (iii) employ mean field-averaging~\cite{SPhysRevA.98.013825,Asgarnezhad_Zorgabad_2020_1,Asgarnezhad-Zorgabad:20_1} to consider the evanescent coupling effect. We note that our waveguide for invariant PFCs generation and SPL formation is based on the source-waveguide-detection triplet. The use of coherent laser fields as a source and multi-level atomic layer as the tunable nonlinear medium has two main advantageous, the PFCs excitation process is coherent and the maximum probe laser peak needed to generate NSPPs is very low. Consequently, NSPP generation is efficient and PFCs propagate through the interaction interface for a few tens of nonlinear propagation length.

Finally, we note that our suggested nonlinear waveguide comprises a multi-level atomic medium situated on top of the ultra-low loss plasmonic layer has two challenges. First, the waveguide is a combination of cold atomic medium, double-layer graphene, and coherent laser sources, and its feasibility in a real-life experiment seems a challenge. Consequently, our specific example contains an explicit formalism to achieve invariant PFCs and SPL formation, but its feasibility is challenging and needs further investigation, whose investigation goes beyond the scope of this current work. Second, invariant PFCs and SPL formation is based on justification of the \textit{conservation conditions}, for which a stable propagation of NSPP provides robust PFC excitation. As a detection system of our specific example, we suggest NSOM~(near-field scanning optical microscope) technique, due to efficiency for collecting output NSPP fields. The suitability of this technique to establish the conservation condition and the efficiency of NSOM technique to invariant PFC generation looks challenging that need further consideration. We leave the suitable detection technique for uncovering invariant PFCs as a future work. 

\section{\label{Sec:Linear_Response}Linear response of a graphene layer}
In this section, we investigate the technical details and mathematical steps towards SPP excitation within our graphene structure. We achieve the loss-compensated plasmonic scheme in three steps: First, we excite the SPP by end-fire coupling of driven laser fields to the graphene-dielectric-graphene multilayer. Next, we suppress the loss related to upper graphene layer by inducing gain to the bottom graphene layer using a photo-inverted scheme introduced in Ref.~\cite{SHess2015_1} for the atomic dipole transition wavelength. The graphene SPP mode for lossy graphene and gain-assisted graphene propagate through this multilayer graphene apparatus. Finally, we couple these two SPP modes using formalism developed in Ref.~\cite{SHwang2007}, to derive the dynamical evolution of the stable graphene SPP mode in the interface between the atomic medium and upper graphene layer.

First, we establish the excitation of SPP wave within graphene layer. This plasmonic field is a TM wave with wavenumber $\bm{k}$, frequency $\omega$, phase $\theta=kx-\omega t$, and field profile
\begin{align}
   \bm{E}(\bm{r},t)=&\bm{E}(z)\exp\{\text{i}\theta\},\label{Eq:Evanescence_Electric}\\
    \bm{H}(\bm{r},t)=&-\bm{e}_{y}H(z)\exp\{\text{i}\theta\}.
    \label{Eq:Evanescence_Magnetic}
\end{align}
This field propagates along atomic medium-graphene layer interface whose current density $\bm{J}=\sigma\bm{E}$ and surface charge density $\rho_\text{ext}$ describe by 
\begin{align}
    \bm{J}:=&\bm{J}_\text{s}\delta(z),\\
    \bm{\rho}_\text{ext}:=&\rho_\text{s}\delta(z).
    \label{Eq:Surface_conditions}
\end{align}
The boundary conditions related to this plasmonic system are
\begin{align}
    \bm{E}_{1\text{t}}=&\bm{E}_{2\text{t}},\;\; \bm{D}_{1\text{n}}-\bm{D}_{2\text{n}}=\rho_\text{s},\label{Eq:Boundry_Conditions_Electric}\\
    \bm{B}_{1\text{n}}=&\bm{B}_{2\text{n}},\;\; \bm{H}_{1\text{t}}-\bm{H}_{2\text{t}}=\bm{J}_\text{s}\times\bm{n}.
    \label{Eq:Boundry_Conditions_Magnetic}
\end{align}
Now, we replace Eqs.~\eqref{Eq:Evanescence_Electric}-\eqref{Eq:Surface_conditions} into Maxwell equations, employ the boundary conditions \eqref{Eq:Boundry_Conditions_Electric}, \eqref{Eq:Boundry_Conditions_Magnetic} to achieve the characteristic dispersion of the SPP wave as
\begin{equation}
    \frac{\varepsilon_{1}}{k_1}+\frac{\varepsilon_{2}}{k_2}-\frac{4\pi  e^2}{k^2} \chi (k,\omega)=0.
    \label{Eq:SPP_single-layer}
\end{equation}
for
\begin{equation}
  \text{i} \omega \chi(k,\omega)=k^2\sigma(k,\omega).
\end{equation}
Finally we define
\begin{equation}
    \tilde{\varepsilon}_{12}:=\left(\varepsilon_{1}^{2}-\varepsilon_{2}^{2}\right)(\varepsilon_{1}\mu_{2}-\varepsilon_{2}\mu_{1}),\;\;\; \tilde{\rho}_\text{s}:=\frac{\rho_\text{s}\omega\varepsilon_{1}\varepsilon_{2}}{C},
\end{equation}
to achieve the propagation constant of the SPP wave
\begin{equation}
    k=\left[\frac{(\omega/\text{c})^{2}\varepsilon_{1}\varepsilon_{2}\tilde{\varepsilon}_{12}+2\tilde{\rho}_\text{s}k\varepsilon_{1}\varepsilon_{2}\sqrt{\tilde{\varepsilon}_{12}+\tilde{\rho}_\text{s}^{2}}+\tilde{\rho}_\text{s}^{2}(\varepsilon_{1}^{2}+\varepsilon_{2}^{2})}{\varepsilon_{1}^{2}-\varepsilon_{2}^{2}}\right]^{1/2},
    \label{Eq:Propagation_Constant}
\end{equation}
here $k_j^{z}$ wavevector component is
\begin{equation}
    k_{j}^{z}=\sqrt{k^{2}-\frac{\omega^{2}\varepsilon_{j}\mu_{j}}{\text{c}^{2}}},\;\;\;\;\; j\in\{1,2\}.
    \label{Eq:SPP_wave_Vector_Simple}
\end{equation}
Eq.~\eqref{Eq:SPP_single-layer} with propagation constant \eqref{Eq:Propagation_Constant} demonstrates the excitation of SPP in our hybrid waveguide~\cite{SBludov2013}. 

Next, we represent the SPP wave excitation within our proposed double layer graphene plasmonic waveguide. Our quantitative approach is for characterizing SPP through this interface is based on (i) random phase~\cite{SWunsch_2006}, and (ii) relaxation time~\cite{SJablan2009,SHess2015_1} approximations. 
Single-graphene layer has valance $\lambda=-$, and conduction $\lambda=+$, bands touching at Dirac points that can be described with chemical potential $\mu$, the interaction interface $A$, spin~(s) and valley~(v) degeneracies $g_{\nu}=2$; $\nu\in\{\text{s},\text{v}\}$, and with electron Fermi velocity $v_\text{F}$~\cite{SNovoselov2012}.
Single layer susceptibility within interaction interface that is obtained using random phase approximation
\begin{equation}
\chi(\boldsymbol{k},\omega)= \frac{g_s g_v}{A} \sum_{q,\lambda , \lambda'=\pm} \frac{n_{q,\lambda}-n_{q+k,\lambda'}}{\hbar \omega+E_{q,\lambda}-E_{q+k,\lambda'}+i0^+}  [1+\lambda \lambda' \cos\left(\theta_\text{q}-\theta_\text{q+k}\right)],
\label{Eq:RPA}
\end{equation}
for $\bm{q}$ the wave vector, $\theta_\text{q}$ the deviation from $x$-axis, and $E_{q,\lambda}=\lambda \hbar v_\text{F} q$ the energy dispersion of electrons with $\lambda$ and $|\bm{q}|$. We evaluate \eqref{Eq:RPA} using Fermi distribution function $n_{q,\lambda}$ for near zero temperatures as $n_{q,\lambda}=\Theta(\mu-E_{q,\lambda})$; with $\mu$ the chemical potential. We calculate \eqref{Eq:RPA} by employing relaxation approximation (RA) as $\tilde{\omega}=\omega+\text{i}\gamma$ to achieve density response for single graphene layer, which includes an undoped part $\chi^0$ and a doped part $\chi^\mu$. This doped part characterises the optical properties of the lossy graphene layer for $\mu\neq0$ through~\cite{SWunsch_2006,Jalaligraphene_1}
\begin{align}
     \chi^{(L)}(k,\tilde{\omega})=&\chi^{(0)}(k,\tilde{\omega})+\chi^{(\mu)}(k,\tilde{\omega}),\\
     \chi^{(0)}(k,\tilde{\omega})=&\frac{-i g_\text{s} g_\text{v} k}{16\hbar \sqrt{\tilde{\omega}^2-v_\text{F}^2 k^2}},\;\;\;
     \chi^{{\mu}}(k,\tilde{\omega})=\frac{g_\text{s} g_\text{v} \mu}{8\pi \hbar^2 v_\text{F}^2}\left[-4+\left(\frac{\hbar v_\text{F} k}{\mu }\right)^2 \frac{G(x^+)+G(x^-)}{2 \sqrt{\tilde{\omega}^2-v_\text{F}^2 k^2} }\right],
    \label{Eq:Graphene-Lossy}
\end{align}
for
\begin{equation}
    G^{\pm}(x)=x \sqrt{x^2-1}-\ln\left(x+\sqrt{x^2-1}\right),\;\;\; x^\pm=\frac{\hbar \tilde\omega\pm2\mu}{\hbar v_\text{F} k},
\end{equation}
and we take into account the collision loss as a relaxation to perturbation frequency by employ mapping
\begin{equation}
    \omega\mapsto\tilde{\omega}=\omega+i \gamma,
    \label{Eq:Relaxation_Time}
\end{equation}
for $\gamma=\tau^{-1}$ the electric relaxation frequency~\cite{SHess2015_1}. Eq.~\eqref{Eq:Graphene-Lossy} commensurate with \eqref{Eq:Relaxation_Time} describes the spectral evolution of the lossy graphene.  Next, we couple a trigger field $I_\text{G}$ to modulate electron~(hole) chemical potential $\mu_\text{e}$~($\mu_\text{h}$) for a gain-assisted graphene medium, and obtain corresponding density response function through mapping
\begin{equation}
    \chi^{(\mu)}(k,\tilde{\omega})\mapsto\chi^{(\mu_\text{e})}(k,\tilde{\omega})+\chi^{(\mu_\text{h})}(k,\tilde{\omega}).
\end{equation}
Next we exploit \eqref{Eq:Graphene-Lossy} to achieve~\cite{SHess2015_1} 
\begin{equation}
    \chi^{(\text{G})}(k,\tilde{\omega})=\chi^{(0)}(k,\tilde{\omega})+\chi^{(\mu_\text{e})}(k,\tilde{\omega})+\chi^{(\mu_\text{h})}(k,\tilde{\omega}).
    \label{Eq:Graphene-Gain}
\end{equation}
Eqs.~\eqref{Eq:Graphene-Lossy} and \eqref{Eq:Graphene-Gain} justify SPP propagation in our hybrid plasmonic system.  

Finally, we investigate the coupling between these gain-loss doublet in our configuration.To this aim,
we assume a dielectric spacer between gain-loss paired graphene layers $\varepsilon_\text{d}$, neglect the orbital overlap and employ Coulomb interaction to achieve the effective susceptibility of the coupled layer as $\bm{\chi^\text{C}}$, and evaluate the hybrid plasmonic system dielectric function as
\begin{equation}
    \bm\varepsilon(k,\omega)=\bm1-\bm\chi^{(\text{C})}(k,\omega) \bm V(k), 
    \label{Eq:Dielectric_Function}
\end{equation} 
in which potential matrix is characterized by diagonal intra-layer $V_{11}=V_{22}$ and off-diagonal inter-layer $V_{12}=V_{21}$ elements~\cite{SProfumo2010}
\begin{align}
    V_{ij}=&\frac{8\pi e^2}{k D}\varepsilon_\text{d},\\
    V_{ii}=&\frac{4\pi e^2}{k D}\left[(\varepsilon_\text{d}+\varepsilon_1)\exp\{k d\}+(\varepsilon_\text{d}-\varepsilon_2)\exp\{-k d\}\right],
    \label{Eq:Potential_Element}
\end{align}
with~\cite{explanation_One}
\begin{equation}
    D=(\varepsilon_1+\varepsilon_d)(\varepsilon_d+\varepsilon_2)\exp\{k d\}+(\varepsilon_1-\varepsilon_d)(\varepsilon_d-\varepsilon_2)\exp\{-k d\}.
    \label{Eq:Denominator}
\end{equation}
In our analysis, we assume the density response of the coupled gain-loss system  $\bm{\chi}^{(\text{C})}(k,\tilde{\omega})$ as a diagonal matrix (i.e. with $\chi^{(\text{C})}_{ij}(k,\tilde{\omega})=0$). Each diagonal element of which represents the corresponding density response of each layer in coupled gain-loss system.
Plugging Eq.~\eqref{Eq:Potential_Element} into \eqref{Eq:Dielectric_Function} would yield the characteristic equation for SPP dispersion~\cite{SHwang2009, jalalidoublelayer_1}
\begin{equation}
    \left[1-V_{11}(k) \chi^{(\text{C})}_{11}(k,\omega) \right]\left[1-V_{22}(k) \chi^{(\text{C})}_{22}(k,\omega) \right]- V_{12}(k) V_{21}(k) \chi^{(\text{C})}_{11}(k,\omega)\chi^{(\text{C})}_{22}(k,\omega)=0,
     \label{Eq:Coupled-Relation}
\end{equation}
which establishes excitation of stable SPP mode for coupled gain-loss graphene layers.
\begin{figure}[t]
    \centering
    \includegraphics[width=0.99\textwidth]{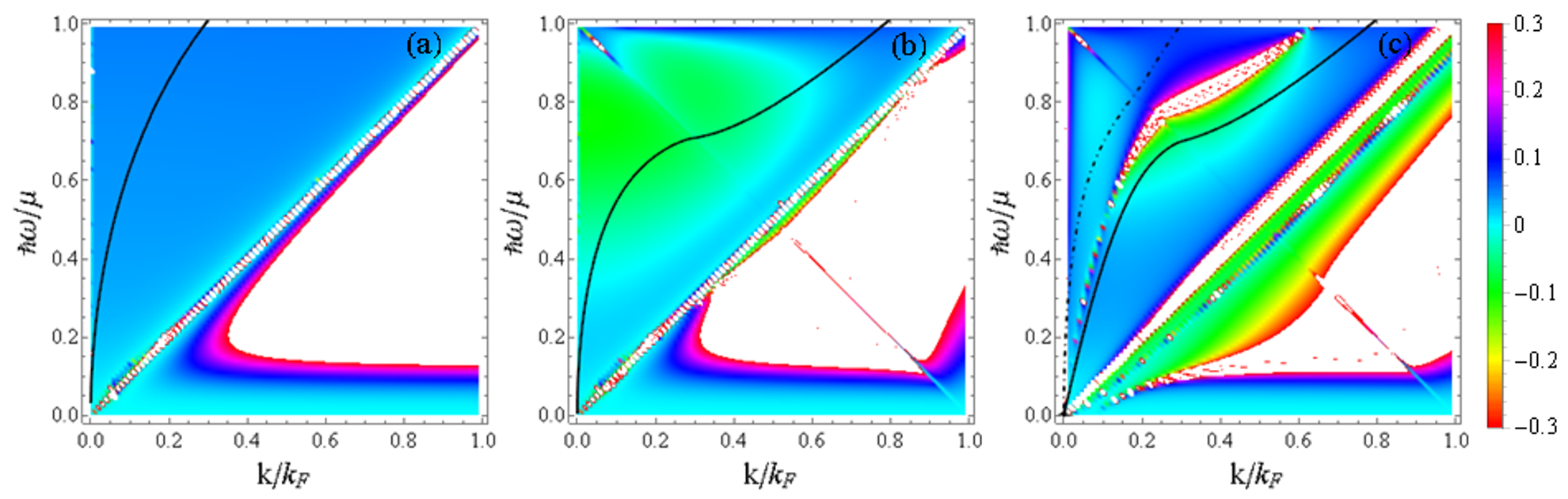}
    \caption{Spectral evolution of the graphene SPP within hybrid interface as a function of normalized energy $\hbar \omega/\mu$ and normalized momentum $k/k_\text{F}$: Panel (a) represent the SPP dispersion for lossy graphene, panel (b) depicts the SPP in a hybrid interface with gain assisted graphene and panel (c) shows the SPP behavior in coupled gain-loss graphene bi-layer. Parameters used for these simulations are: $\epsilon_1=\epsilon_2=1$, $g_\text{s}=g_\text{v}=2$, $\text{c}=3\times10^8~\text{m}/\text{s}$ and $v_\text{F}=10^6~\text{m}/\text{s}$,  $\hbar\tau^{-1}/\mu=0.08 \text{Hz}$~\cite{SHess2015_1}. For panel (c) $\varepsilon_\text{d}\approx2$ and $k_\text{F}d=4.43$~\cite{SHwang2009}. Normalized waveguide decay is characterized by $\hbar \gamma/\mu$ as a color bar. See text for detailed explanation.}
  \refstepcounter{SIfig}\label{Fig:Fig_graphene_lossy}
\end{figure}

We represent the spectral evolution of the excited SPPs for single lossy graphene, gain assisted graphene and gain-loss paired double graphene layers in Fig.~\ref{Fig:Fig_graphene_lossy}. The SPP within Pauli-blocked inter-band characterized by $\hbar(\omega + v_\text{F}k)<2\mu$ and $\omega>v_\text{F}k$ would be dissipative along graphene-dielectric interface due to relaxation decays as clearly shown in Fig.~\ref{Fig:Fig_graphene_lossy}(a). For a gain assisted graphene, however, we couple a trigger laser to induce population inversion between valance and conduction band, hence the gain is provided for effective zero carrier density $\mu_\text{h}=\mu_\text{e}=\mu/2$~\cite{SHess2015_1}. This laser-induced mechanism yields modification in graphene density response, suppresses the loss due to additional gain induction and consequently results in loss-free propagation of graphene SPP as we establish in Fig.~\ref{Fig:Fig_graphene_lossy}(b). Finally, in a gain-loss paired graphene double layers separated by a spacer, two-mode graphene SPP propagation is expected due to coupled-mode theory~\cite{SHwang2009}, which are propagated along the hybrid interface and characterized by dashed-dotted and solid lines in Fig.~\ref{Fig:Fig_graphene_lossy}(c). We consider the dynamical evolution of the solid line SPP mode due to gain-assisted loss compensation. The dashed-dotted line SPP mode is highly dissipative and consequently its propagation length is highly limited due to total loss of graphene layer.    

We test the feasibility of loss-free SPP mode propagation along gain-loss paired double-layers by choosing the realistic parameters. The carrier density is~\cite{SFerrari2010graphene}
\begin{equation}
    n_\text{s}=\frac{g_\text{s} g_\text{v} \mu^2}{4\pi\hbar^2  v_\text{F}^2}\approx10^{11}~\text{cm}^{-2},
\end{equation}
and the Fermi surface momentum is $k_\text{F}=10^9 \text{m}^{-1}$. Considering the degeneracies as $g_{\nu}=2$, and $v_\text{F}=c/300$, the low-loss propagation of the SPP within this coupled plasmonic scheme would be excited for our dipole transition wavelength $\lambda=800~\text{nm}$ and $k/k_\text{F}\approx0.5$~\cite{SNovoselov2012}, which establishes our assumptions within Fig.~\ref{Fig:Fig_graphene_lossy}(c). Consequently, this SPP mode interacts with the nonlinearity and dispersion of the interface and possess spatiotemporal evolution.

\section{\label{SM:Invariant_evolution} Spectral evolution of the plasmonic frequency combs in the presence of invariants}
In this section, we provide the main steps towards the mathematical details of the derivation of the Eq.~\eqref{Eq:Modified_NLSE}. We present our quantitative approach in two subsections. First in \S~\ref{Sec:Existencce_Invariant}, we use a classical treatment and employ nonlinear modification to establish the existence of system invariant through our nonlinear dissipative interface. Next, in \S~\ref{Eq:Quantum_Averaging} we employ the quantum theory of soliton~\cite{SPhysRevA.40.844} and employ mean-value quantum field evolution to achieve the spectral field dynamics of the plasmonic frequency combs in the presence of conserved energy and conserved number of excited SPP modes, thereby derive Eq.~\eqref{Eq:Modified_NLSE} of the main text. 

\subsection{\label{Sec:Existencce_Invariant}Existence of system conservatives and derivation of Eq.~\eqref{Eq:Invariant_Nonlinear}}
We represent the qualitative approach towards system conservatives in two sub-sections. First, we evaluate the dynamical evolution ot the nonlinear SPP field within our interaction interface and derive Eq.~\eqref{Eq:Dynamics_SPP_Field} of the main text. Next, we introduce energy and number of excited SPP mode as the two nonlinear parameters that affect the dynamics of frequency combs and establish the conservations of these quantities through nonlinear interaction, thereby present detailed derivation of Eq.~\eqref{Eq:Invariant_Nonlinear}.

\subsubsection{Nonlinear SPP field dynamics and derivation of Eq.~\eqref{Eq:Dynamics_SPP_Field}}
Our starting point is the propagation of the field within the nonlinear interface
\begin{equation}
    \bm{\nabla}\times\bm{\nabla}\times\bm{E}+\frac{1}{\text{c}^2}\frac{\partial^2\bm{D}}{\partial t^2}=-\frac{1}{\varepsilon_0\text{c}^2}\frac{\partial^2\bm{P}}{\partial t^2},
    \label{Eq:Maxwel_Wave}
\end{equation}
for
\begin{align}
    \bm{D}(\bm{r},t)=&\int_0^{t_\text{S}}\text{d}t'\varepsilon(t')\bm{E}(\bm{r},t-t'),\label{Eq:Dielectric_Linear}\\ \bm{P}(\bm{r},t)=&\bm{E}(\bm{r},t)\int_0^{t_\text{S}}\text{d}t'\chi^{(3)}(\bm{r},t')|\bm{E}(\bm{r},t-t')|^2,\label{Eq:Nonlinear_Polarizaion}
\end{align}
the displacement vector and nonlinear polarization, respectively and for $\varepsilon(t')$, $\chi^{(3)}(t')$ the susceptibility and Kerr nonlinear coefficient of the interaction interface. In the Fourier space, we represent the electric field as $\Psi(\bm{r},\omega)$ with
\begin{equation}
    \Psi(\bm{r},\omega):=F(\bm{r},\omega)\tilde{A}(x,\omega),
    \label{Eq:SPP_Field}
\end{equation}
$F(\bm{r},\omega)$ the spectral-spatial mode distribution in the interaction interface, $\tilde{A}(x,\omega)$ the field distribution along interaction direction and $\mathcal{K}(\tilde{\omega})$ the dispersion of the plasmonic field. We assume the Fourier transform of the probe field as
\begin{equation}
    E(\bm{r},t)=\int\;\text{d}\omega\Psi(\bm{r},\omega)\exp\{\text{i}\mathcal{K}(\omega)x-\text{i}\omega t\}.
    \label{Eq.Fourier_Integral}
\end{equation}

We exploit Eq.~\eqref{Eq.Fourier_Integral} to evaluate temporal dynamics of the displacement vector and nonlinear polarization as
\begin{align}
    \frac{\partial^2\bm{D}}{\partial t^2}=&-\int\;\text{d}\omega \left[\omega^2\varepsilon(\bm{r},\omega)F(\bm{r},\omega)\tilde{A}(x,\omega)\exp\{\text{i}\mathcal{K}x\}\right],\label{Eq:Dielectric_Linear_Fourier}\\
    \frac{\partial^2\bm{P}}{\partial t^2}=&-\int\;\text{d}\omega \left[\omega^2\iint\;\text{d}\omega'\text{d}\nu\exp\{\text{i}\Delta\mathcal{K}x\}\chi_{\omega-\nu}^{(3)}(\bm{r})\tilde{A}_{\omega-\nu+\omega'}(x)F_{\omega-\nu+\omega'}(r)\tilde{A}_{\nu}^{*}(x)F_{\nu}^{*}(r)\tilde{A}_{\omega'}(x)F_{\omega'}(r)\right],\label{Eq:Nonlinear_Polarizaion_Fourier}
\end{align}
with
\begin{equation}
    \Delta\mathcal{K}=\mathcal{K}(\omega-\nu+\omega')-\mathcal{K}(\nu)+\mathcal{K}(\omega')-\mathcal{K}(\omega),
\end{equation}
and we also ignore the spatial distribution of the optical Kerr nonlinearity $\chi_{\omega-\nu}^{(3)}(\bm{r})\approx\chi^{(3)}(\omega-\nu)$. Next, we substitute Eqs.~\eqref{Eq:Dielectric_Linear_Fourier} and \eqref{Eq:Nonlinear_Polarizaion_Fourier} into Eq.~\eqref{Eq:Maxwel_Wave} and exploit
\begin{equation}
    \mathcal{K}^2=\frac{\varepsilon^2(\bm{r},\omega)\omega^2}{\text{c}^2}
\end{equation}
to achieve
\begin{align}
    \bm{E}(x,\omega)\nabla_{\perp}^2F(\bm{r},\omega)+F(\bm{r},\omega)\left[2\text{i}\mathcal{K}(\omega)\frac{\partial}{\partial x}+\frac{\partial^2}{\partial x^2}\right]\bm{E}(\bm{r},t)=\nonumber\\-\frac{\omega^2}{\varepsilon_0^2\text{c}^2}\iint\;\text{d}\omega'\text{d}\nu\exp\{\text{i}\Delta\mathcal{K}x\}\chi_{\omega-\nu}^{(3)}(\bm{r})\tilde{A}_{\omega-\nu+\omega'}(x)F_{\omega-\nu+\omega'}(r)\tilde{A}_{\nu}^{*}(x)F_{\nu}^{*}(r)\tilde{A}_{\omega'}(x)F_{\omega'}(r)&.
    \label{Eq:First_Spectrum_Dynamics}
\end{align}
In this work we employ slowly varying amplitude approximation~(i.e. $\partial^2/\partial x^2\mapsto0$). We then multiple both sides of Eq.~\eqref{Eq:First_Spectrum_Dynamics} by $F(\bm{r},\omega)$ and perform integration over all possible transverse coordinates. Using 
\begin{equation}
    \bm{E}(x,\omega)\int\;\text{d}\bm{r}F(\bm{r},\omega)\nabla^2F(\bm{r},\omega)=0,
\end{equation}
define the Green's function of the medium as
\begin{equation}
    G(\omega,\omega',\nu):=\frac{\int\;\text{d}\bm{r}F_{\omega}(r)F_{\omega'}(r)F_{\nu}^{*}(r)F_{\omega-\nu+\omega'}(r)}{\int\;\text{d}\bm{r}F_{\omega}^{2}(r)},
    \label{Eq:Green_Function}
\end{equation}
and effective refractive index of the medium as
\begin{equation}
    n_\text{eff}(\omega)=\frac{\text{c}\mathcal{K}(\omega)}{\omega}
\end{equation}
we achieve the spectral evolution of the field in the interaction interface as
\begin{equation}
    \frac{\partial \Psi(\omega,x)}{\partial x}=\frac{2\text{i}\pi\omega}{n_\text{eff}(\omega)\text{c}}\iint\;\text{d}\omega'\text{d}\nu\chi^{(3)}(\omega-\nu)G(\omega,\omega',\nu)\tilde{A}_{\omega'}(x)\tilde{A}_{\nu}^{*}(x)\tilde{A}_{\omega-\nu+\omega'}(x).
\end{equation}

In this work, we neglect the field variation along longitudinal direction $y$ and we assume the field is concentrate at interface through the transverse direction as $|\bm{E}(z)|\sim\exp\{-\text{Im}[\mathcal{K}(\omega)]z\}$. We assume this coupling coefficient as $\zeta(z)$. Our predicted nonlinear field propagation, therefore, is valid for effective propagation length $L_\text{eff}$ and interaction interface $S_\text{eff}$ that can be evaluated using the transverse distribution and Eq.~\eqref{Eq:Green_Function} as
\begin{align}
    L_\text{eff}=&\frac{1}{\alpha_\text{eff}}\left[1-\exp\{-a_\text{eff}x_\text{max}\}\right],\\
    S_\text{eff}=&\frac{\left(\int_{-\infty}^{+\infty}\int_{-\infty}^{+\infty}\;\text{d}x\text{d}z\left|\zeta(z)F(y,z;\omega')\right|^{2}\right)^2}{\int_{-\infty}^{+\infty}\int_{-\infty}^{+\infty}\;\text{d}x\text{d}z\left|\zeta(z)F(y,z;\omega')\right|^4}.
\end{align}
Now we define the coefficient characterizing the self-phase modulation $\mathcal{W}(\omega)$ as
\begin{equation}
    \mathcal{W}(\omega'):=\frac{n_2(\omega')}{\text{c}S_\text{eff}}.
    \label{Eq:SPM_Coeffficient}
\end{equation}
In our analysis, the hybrid interface possesses ultra-low Ohmic loss only for the small deviation of the SPP field frequency. as a result, consiedring the small frequency perturbation as $\omega'=\omega+\omega_\text{SPP}$, the self-phase modulation can be expanded as a Tylor series
\begin{equation}
    \mathcal{W}(\omega)=\mathcal{W}_0+\mathcal{W}_1\delta\omega+\mathcal{W}_2\delta\omega^2+\mathcal{O}(\delta\omega^3).
    \label{Eq:Nonlinear_Coefficient}
\end{equation}
Our predicted nonlinear plasmonic effects are valid for certain coherence timescale that we evaluate by plugging Eq.~\eqref{Eq:Nonlinear_Coefficient} to Eq.~\eqref{Eq:SPM_Coeffficient} and truncate the Taylor expansion only to first order. The specific nonlinear timescale then is
\begin{equation}
    t_\text{S}=\tau_0+\frac{\text{d}}{\text{d}\omega}\left[\ln\left(\frac{1}{n_\text{eff}S_\text{eff}}\right)\right]_{\omega=\omega_\text{SPP}},
\end{equation}
that is defined as the characteristic time-length for which the nonlinear interaction of the system can be described by the nonlinearity modulated as Eq.~\eqref{Eq:Nonlinear_Coefficient}. 

For the modulated SPP field characterised as Eq.~\eqref{Eq:SPP_Field}, we achieve the propagation constant as $\mathcal{K}(\omega):=\beta(\omega)+k(\omega)$; $\beta(\omega)$ is the linear chromatic dispersion of the atomic medium and $k(\omega)$ given by Eq.~\eqref{Eq:SPP_Field}. Consequently, we include this dispersion into the spectral evolution of the SPP field to achieve
\begin{equation}
    \frac{\partial \Psi(\omega,x)}{\partial x}=\text{i}\mathcal{K}(\omega)\tilde{A}(x,\omega)+\frac{2\text{i}\pi\omega}{n_\text{eff}(\omega)\text{c}}\iint\;\text{d}\omega'\text{d}\nu\chi^{(3)}(\omega-\nu)G(\omega,\omega',\nu)\tilde{A}_{\omega'}(x)\tilde{A}_{\nu}^{*}(x)\tilde{A}_{\omega-\nu+\omega'}(x).
    \label{Eq:Initial_Nonlinear_Spectral}
\end{equation}
Eq.~\eqref{Eq:Initial_Nonlinear_Spectral} contains a nonlinear term that acts as a convolution that connects the amplitudes of the SPP field with different amplitudes. This term hence characterizes the nonlinear interaction through the interface. In this work, we aim to investigate the plasmonic frequency combs, which connects the three nearest neighbor frequencies $\omega-\omega_m$, $\omega$ and $\omega+\omega_m$ and we assume the frequency combs are equally distanced and descritized. Therefore the frequency indices in Eq.~\eqref{Eq:Initial_Nonlinear_Spectral} should change to $\omega$, $\omega\pm\omega_m$, respectively and we also consider mapping $\int\;\text{d}\nu\mapsto\sum_m$ to includes all the stable frequency combs. Then we achieve
\begin{equation}
   \text{i}\frac{\partial\tilde{\Psi}(\bm{r},\omega)}{\partial x}=\mathcal{K}(\omega)\tilde{\Psi}+\iint\;\text{d}\nu\text{d}\omega'\frac{\mathcal{W}^0(\omega')}{(2\pi)^2}\tilde{\Psi}_{\omega'-\nu}^{*}(\bm{r},\omega')\tilde{\Psi}_{\omega'}(\bm{r},\varpi)\tilde{\Psi}_{\omega'+\nu}(\bm{r},\nu)\text{e}^{\text{i}\Delta\mathcal{K} x},
    \label{Eq:NLSE_Spatiotemporal_First}
\end{equation}
that is Eq.~\eqref{Eq:Dynamics_SPP_Field} of the main text.

\subsubsection{\label{Sec:Invariants_NSPP}Existence of conservation and nonlinear plasmonic field modulation}
In this section, we elucidate our quantitative description towards existence of conserved parameters and then modulate the nonlinearity for simultaneous conservation of energy and number of excited SPP mode. We notice that the frequency combs then excite due to nonlinearity and dispersion management similar to Ref.~\cite{SPhysRevA.99.051802}. These plasmonic combs would possess ultra-low loss of the frequencies within electromagnetically induced transparency windows of the atomic medium due to suppressed dissipation. In this interface with modulated nonlinearity characterized by Eq.~\eqref{Eq:Nonlinear_Coefficient} these combs take the form
\begin{equation}
    \Psi(\bm{r},t)\sim\sum_{m=1}^{N_\text{EIT}}A_m(x)\exp\{\text{i}\omega_m t\},
\end{equation}
the total energy of the excited SPP frequency combs are
\begin{equation}
    E\propto\sum_{m=1}^{N_\text{EIT}}\left|A_m(x)\right|^2,
    \label{Eq:Energy_NonlinearSystem}
\end{equation}
and we take the number of excited frequency combs as
\begin{equation}
    \mathcal{N}\propto\sum_{m=1}^{N_\text{EIT}}\frac{\left|A_m(x)\right|^2}{\omega_\text{SPP}+\omega_m}.
    \label{Eq:NumberModdes_NonlinearSystem}
\end{equation}
We then take derivative with respect to $x$ from both side of Eqs.~\eqref{Eq:Energy_NonlinearSystem} and \eqref{Eq:NumberModdes_NonlinearSystem} and evaluate the spatial variation of the energy $\partial E/\partial x$ and number of excited frequency combs $\partial \mathcal{N}/\partial x$ as
\begin{align}
        \frac{\partial E}{\partial x}\propto&-\bar{\alpha}\sum_{m}|A_m(x,\tilde{\omega})|^2+\sum_{m}[\mathcal{W}(\omega_\text{ch})+\mathcal{W}(\omega_{0})-\mathcal{W}(\omega_{-})-\mathcal{W}(\omega_{+})]\Delta,\label{Eq:Dynamics_Energy}\\
        \frac{\partial\mathcal{N}}{\partial x}\propto&-\bar{\alpha}\sum_{m}\frac{|A_m(x,\tilde{\omega})|^2}{\omega_0+\omega_m}+\sum_{m}\left[\frac{\mathcal{W}(\omega_{0})}{2\omega_0}+\frac{\mathcal{W}(\omega_{-})}{\omega_0+\omega_{-}}+\frac{\mathcal{W}(\omega_{+})}{\omega_0+\omega_{+}}\right]\Delta,\label{Eq:Dynamics_Number}
\end{align}
for $\mathcal{W}(\omega_\text{ch})/(\omega+\omega_\text{ch})\approx0$. In writing Eqs.~\eqref{Eq:Dynamics_Energy} and \eqref{Eq:Dynamics_Number} we assume
\begin{equation}
    \Delta:=4\text{Im}\left[A_{1}^{*}A_2A_3A_{4}^{*}\exp\{\text{i}\Delta\mathcal{K}_\text{t}x\}\right]
\end{equation}
as the detuning of the SPP fields through four-wave mixing process with
\begin{equation}
    \Delta\mathcal{K}_\text{t}=\mathcal{K}(\omega_\text{ch})+\mathcal{K}(\omega_{+})-\mathcal{K}(\omega_{0})-\mathcal{K}_4(\omega_{-}),
\end{equation}
denotes the phase mismatch between the different SPP field and
\begin{equation}
    \mathcal{K}(\omega_{l})=\beta(\omega_{l})+k(\omega_l)+\sum'\sum_{n}\left(2-\delta_{ln}\right)\mathcal{W}(\omega_{l})|A_n(\omega_l)|^2,
    \label{Eq:Nonlinear_Wavenumber}
\end{equation}
for $n\in\{+,-,0,\text{ch}\}$, represent the nonlinear wavenumber of the plasmonic interface. Also, the prime in Eq.~\eqref{Eq:Nonlinear_Wavenumber} denotes that the summation performed over all possible frequency combs.

It is obvious from Eq.~\eqref{Eq:Dynamics_Energy} that by modulating $\mathcal{W}(\omega)=\mathcal{W}_0+\mathcal{W}_1\omega$ the energy becomes invariance of the system but the number of excited frequency combs would be varying. On the other hand, by choosing $\mathcal{W}(\omega)=\mathcal{W}'_1(\omega+\omega_\text{SPP})+\mathcal{W}'_2(\omega+\omega_\text{SPP})^2$, the number of modes become the invariant off the system, but we loose the energy conservation. However, the simultaneous invariant for both energy and number of excited SPP waves is achieved through nonlinearity modulation as
\begin{equation}
    \mathcal{W}=\mathcal{W}_0+\left(\frac{\mathcal{W}_0}{\omega_\text{SPP}}\right)\omega,
    \label{Eq:Modificed_Nonlinearity_System}
\end{equation}
that is Eq.~(\textcolor{blue}{5}) of the main text for $\omega_\text{SPP}\mapsto\omega_0$. As it is clear from Eq.~\eqref{Eq:Modificed_Nonlinearity_System}, energy and number of excited SPP modes would become conserved parameter of the system if we linearize the nonlinearity as this equation and also perturb the frequency only for $\tilde{\omega}=\omega+\omega_\text{SPP}$. We refer these two conditions as the conservation conditions for the plasmonic system.

\subsection{\label{Eq:Quantum_Averaging}Mean-field evolution of quantum nonlinear SPP mode and derivation of Eq.~(\textcolor{blue}{6})}
In this section we elucidate our quantitative approach for the derivation of Eq.~Eq.~(\textcolor{blue}{6}) of the main text. We notice that our conservative parameters do not depend on the dispersion of the system, then we can set $\mathcal{K}(\omega)=0$ To obtain this equation, we use the quantum theory approach of optical soliton within a nonlinear fiber~\cite{SPhysRevA.40.844} and extend it to our dissipative hybrid nonlinear interface. Similar to Ref.~\cite{SPhysRevA.40.844}, we assume that our nonlinear Schr\"odinger equation can also be derived through mean-value evolution of the quantum SPP field operator through Sch\"odinger equation
\begin{equation}
    \frac{\partial}{\partial x}\ket{\psi}=H_\text{I}\ket{\psi}.
    \label{Eq:Schrodinger}
\end{equation}
We introduce $\ket{\psi}$ as the quantum state of light, and
\begin{equation}
    \mathcal{H}_\text{I}=\sum_{m}\iint\;\text{d}\tilde{\omega}\text{d}\omega\frac{\mathcal{W}^0}{2}\hat{\bm{b}}_{\omega}^{\dagger}\hat{\bm{b}}_{\tilde{\omega}}^{\dagger}\hat{\bm{b}}_{\tilde{\omega}-\omega_m}\hat{\bm{b}}_{\tilde{\omega}+\omega_m}.
    \label{Eq:Appendix_Interaction_Hamiltonian}
\end{equation}  
as the nonlinear interaction Hamiltonian of the SPP field. 

Here we assume the most general case for which $\hat{\bm{b}}_{\jmath}(\bm{r},\omega)$~($\hat{\bm{b}}_{\jmath}^{\dagger}(\bm{r},\omega))$; $\jmath\in\{\text{e},\text{m}\}$ as annihilation~(creation) operators associated with the electrical~(e) and magnetic~(m) response of the medium, whose components are described by usual bosonic commutation relation
\begin{align}
    \left[\hat{b}_{\jmath i}(\bm{r},\omega),\hat{b}_{\jmath' j}(\bm{r}',\omega')\right]=&0,\label{Eq:Commutation_Relation_First}\\
    \left[\hat{b}_{\jmath i}(\bm{r},\omega),\hat{b}_{\jmath' j}^{\dagger}(\bm{r}',\omega')\right]=&\delta_{ij}\delta_{\jmath\jmath'}\delta(\omega-\omega')\delta(\bm{r}-\bm{r}'). \label{Eq:Commutation_Relation_Second}
\end{align}
$\mathcal{W}^0$ contains the frequency dependent, but we assume the frequencies are all exist within the stable NSPP frequency range. In this case, Eq.~\eqref{Eq:Appendix_Interaction_Hamiltonian} yields stable multiple plasmonic four-wave mixing. As in \S~\ref{Sec:Existencce_Invariant} to consider the energy and number of excited frequency combs as conservatives of the system, we employ mapping $\omega\mapsto\omega_m+\omega_\text{SPP}$, and assume $\mathcal{W}^)=\mathcal{W}^{0*}$. The quantum SPP field within the interaction interface takes the form
\begin{equation}
    \bm{\Psi}(\bm{r},t)=\int_{\bm{r}',\tilde{\omega}}\hbar\tilde{\omega}\left[\bm{\mathcal{A}}(\bm{r},\bm{r}';\tilde{\omega})\cdot\hat{\bm{j}}(\bm{r}',\tilde{\omega})\text{e}^{\text{i}\tilde{\omega}t}+\text{h}.\text{c}\right].
    \label{Eq:Quantized_Electric_Field_Text}
\end{equation}
for $\bm{\mathcal{A}}(\bm{r},\bm{r}';\tilde{\omega})$ the green function of the interface~\cite{Sasgarnezhad2021coherent},
\begin{equation}
    \hat{\bm{j}}(\bm{r},\omega)=-2\pi\text{i}\omega\alpha(\bm{r},\omega)\hat{\bm{b}}_\text{e}(\bm{r},\omega),
    \label{Eq:Quantized_Current_Density}
\end{equation}
the quantized current density of the graphene interface and with
\begin{equation}
    \alpha(\bm{r},\omega)=\left\{\frac{\hbar\varepsilon_0}{\pi}\text{Im}[\varepsilon(\bm{r},\omega)]\right\}^{1/2},\label{Eq:Definition_Permittivity}
\end{equation}
the constant of the system depends on the medium. Plugging Eqs.~\eqref{Eq:Quantized_Current_Density} and~\eqref{Eq:Definition_Permittivity} into Eq.~\eqref{Eq:Quantized_Electric_Field_Text} and making use of \eqref{Eq:SPP_Field}, we see that the quantized plasmonic frequency combs $\hat{A}_{m}$ are
\begin{equation}
    \hat{A}_{m}(\bm{r},\omega)\propto\sqrt{\frac{\hbar\omega}{\varepsilon_0 V}}\hat{\bm{b}}(\bm{r},\omega)+\text{c.c.}.
    \label{Eq:Normalized_Field}
\end{equation}
Next, we substitute the necessary condition for energy and number of SPP mode conservation (i.e. $\omega\mapsto\omega_m+\omega_\text{SPP}$) and substitute into Eq.~\eqref{Eq:Schrodinger} we achieve the mean value evolution of the stable plasmonic frequency combs in the presence of the conservatives of the system as
\begin{equation}
    \frac{\partial \braket{\hat{A}_m}}{\partial x}=\text{i}\iint\;\text{d}\tilde{\omega}\text{d}\omega\Lambda(\omega,\tilde{\omega},\omega_m)\hat{\bm{b}}_{\omega}^{\dagger}\hat{\bm{b}}_{\tilde{\omega}}^{\dagger}\hat{\bm{b}}_{\tilde{\omega}-\omega_m}\hat{\bm{b}}_{\tilde{\omega}+\omega_m}.
    \label{Eq:Field_Evolution}
\end{equation}
$\Lambda$ the nonlinearity of four-wave mixing process for frequency around $\omega_\text{SPP}$, which we define as
\begin{equation}
    \Lambda\propto\left\langle\frac{(\mathcal{W}^0_{\omega,\tilde{\omega},\omega-\omega_m,\tilde{\omega}+\omega_m}+\mathcal{W}^0_{\tilde{\omega},\omega,\tilde{\omega}+\omega_m,\omega-\omega_m})\sqrt{\hbar(\omega+\omega_\text{SPP})/\varepsilon_0V}}{2\sqrt{(\omega_\text{SPP}+\omega-\omega_m)(\omega_\text{SPP}+\tilde{\omega}+\omega_m)(\omega_\text{SPP}+\tilde{\omega})}}\right\rangle_{z},
    \label{Eq:Nonlinear_Modified}
\end{equation}
and we consider the SPP field properties as evanescence coupling and employ field-averaging as in Ref.~\cite{SPhysRevA.98.013825}. 

Eq.~\eqref{Eq:Nonlinear_Modified} is obtained from mean-field evolution of the quantum plasmonic frequency combs in the presence of the SPP field dispersion and dissipation and represents the nonlinearity, which is obtained in the presence of energy and number of excited SPP modes conservation. Due to quantum theory of soliton~\cite{SPhysRevA.40.844}, we expect the nonlinearity is equivalent to the nonlinearity obtained in NLSE. Therefore, in order to modulates the nonlinearity to include conservatives to NLSE, we employ mapping $\Lambda\mapsto\mathcal{W}$ in Eq.~\eqref{Eq:NLSE_Spatiotemporal_First}. Next, we include the dispersion of the system as $\mathcal{K}(\omega)$. Finally we define
\begin{equation}
    \iint\;\text{d}\omega\text{d}\tilde{\omega}\mapsto\mathcal{F}
\end{equation}
as Fourier transform operator. In this case by substituting into Eq.~\eqref{Eq:Field_Evolution} we achieve
\begin{equation}
    \frac{\partial\tilde{A}}{\partial x}=\text{i}\mathcal{K}(\omega)\tilde{A}+\sum_m\mathcal{F}\left[\Lambda(\tilde{\omega})|A_m|^{2}A_m\right]+\text{c.c.},
    \label{Eq:Modified_NLSE_appendix}
\end{equation}
which is the Eq.~(\textcolor{blue}{6}) of the main text. This equation is the same as nonlinear Sch\"odinger equation, as it includes the nonlinear parameters of the system, however this equation differs trivial nonlinear Schr\"odinger equation as we introduce the energy and number of excited SPP modes as the conservatives of the system.

\section{\label{Sec:Construction_synthetic}Construction of plasmonic synthetic lattice}
In this section we present the detailed steps towards synthetic lattice formation within our hybrid nonlinear plasmonic interface. Our quantitative method for constructing the synthetic lattice is based on two steps. First, we explore the NSPP dynamics and excitation of the NSPPs within hybrid interface in \S~\ref{Sec:Formation_NSPP}, and next, we elucidate the main steps towards mapping to a synthetic lattice in \S~\ref{Sec:Formation_Synthetic_Lattice}. It is worth noting to indicate that the formation of this synthetic lattice is crucially depend on the existence of invariants of the NSPPs. Formation of the synthetic lattice corresponds to lossy NSPPs needs further consideration and can be considered as a future work.  

\subsection{\label{Sec:Formation_NSPP} Formation of NSPP and generating frequency combs through interaction interface}
In this section, we elucidate the main steps towards NSPPs excitation in normalized length~($\xi:=x/L_\text{NL}$) and normalized time~($\varrho:=\tau/\tau_0$) and then establish the existence of surface polaritonic frequency combs. The dynamical evolution of a normalized SPP field~($u(\xi,\varrho):=\Omega_\text{p}(\xi,\varrho)/U_0$) in the interface between a nonlinear medium, and a metallic-like interface, whose nonlinear parameters are characterized by self-phase modulation~($W$) and group-velocity dispersion~($K_{2}$), is described by nonlinear Schr\"odinger equation~\cite{SPhysRevA.98.013825,SPhysRevA.99.051802}
\begin{equation}
    \text{i}\frac{\partial u(\xi,\varrho)}{\partial\xi}+\frac{1}{2}K_2\frac{\partial^2u(\xi,\varrho)}{\partial\varrho^2}+W|u(\xi,\varrho)|^2u(\xi,\varrho)=0.
    \label{Eq:NLSE_Spatiotemporal}
\end{equation}

Similar to other nonlinear systems, Eq.~\eqref{Eq:NLSE_Spatiotemporal} possesses exact solutions that are known as peregrine waves $u_\text{Pe}(\xi,\varrho)$ and Akhemediev breather $u_\text{AB}(\xi,\varrho)$~\cite{Skibler2010peregrine}. Next, we employ a Fourier transform of these exact solutions
\begin{equation}
    A_{m}(\xi,\tilde{\omega})=\frac{1}{\sqrt{2\pi}}\int_{0}^{\infty}\;\text{d}\varrho u_{l}(\xi,\varrho)\exp\{\text{i}\tilde{\omega}_m\varrho\},\quad l\in\{\text{Pe},\text{AB}\}, 
\end{equation}
and evaluate the integral to obtain the spectral harmonic side-band amplitudes $A_m(\xi)$ for $\tilde{\omega}=\omega\pm m\Omega/2$; $m\in\{\pm2,\pm4,\ldots\}$ and we define $\Omega:=\omega_\text{EIT}/\mathcal{N}$ as frequency spacing. In our plasmonic interface, we achieve the harmonic side-bands amplitudes correspond to resonant mode $A_0$ as and other frequency side-band as
\begin{equation}
    A_0(\xi)=1-\frac{\text{i}b\sinh\{b\xi\}+p^2\cosh\{b\xi\}}{\sqrt{\cosh^2\{b\xi\}}-2a},
    \label{Eq:Zeroth_sidebandAB}
\end{equation}
and higher-order spectral side-band as
\begin{equation}
    A_m(\xi)=\frac{\text{i}b\sinh\{b\xi\}+p^2\cosh\{b\xi\}}{\sqrt{\cosh^2\{b\xi\}}-2a}\left[1-\frac{\cosh\{b\xi\}{\sqrt{\cosh^2\{b\xi\}}-2a}}{\sqrt{2a}}\right]^{|m|}.
    \label{Eq:Higher_Sideband_AB}    
\end{equation}
Here, we assume $P_0$ as the input power of the SPP field, and also consider $\omega$ as the modulation frequency off the SPP field. Also, we define
\begin{equation}
    \omega_\mathcal{C}=\sqrt{\frac{4\gamma P_0}{|K_2|}},
\end{equation}
as the characteristic frequency of the system. Then we achieve the modulation parameters $a,b$ in terms of these quantities as
\begin{equation}
    2a=\sqrt{1-\left(\frac{\omega}{\omega_\mathcal{C}}\right)^2},\quad b=\sqrt{8a(1-2a)},
\end{equation}
In the limiting case $a\mapsto0.5$ SPP will propagated as plasmonic peregrine wave and we find the spectral harmonic side-band amplitudes by employing harmonic $a\mapsto0.5$ to Eqs.~\eqref{Eq:Zeroth_sidebandAB},~\eqref{Eq:Higher_Sideband_AB}.

The spatial-spectral evolution of these harmonic side-bands are represented in Fig.~\ref{Fig:Fig_Sone_SM}(a). Consequently, generation of NSPPs within a nonlinear hybrid plasmonic interface would yield the excitation of harmonic side-bands with characteristic frequency $\omega_m$ and amplitude $A_{m}(\xi)$ that are propagated along the interaction interface up to a few nonlinear propagation length.    

\subsection{\label{Sec:Formation_Synthetic_Lattice}Synthetic lattice formation based on NSPP parameters}
In this section we elucidate the necessary steps towards synthetic lattice formation of propagated frequency combs. To this aim, first we describe the general properties of the synthetic lattice and describe the sites and hopping related to this system, and next we explain the synthetic lattice Hamiltonian and elucidate the evolution of the synthetic structure in terms of system parameters.

\subsubsection{\label{Sub_Sec:General_Prop}General properties}
In this section we elucidate the detailed steps towards SPL formation based on excited plasmonic frequency combs and characterize the hopping in terms of nonlinear SPP field parameter. Qualitatively, the excited frequency combs within hybrid plasmonic interface are harmonic side-bands with frequency spacing $\Omega$ that act as sites of the lattice. The specific lattice lattice sites are connected to other side-bands through characteristic hoppings~($u_i$) that can be characterized through correlation between amplitudes of spectral harmonic frequencies.

The characteristic lattice sites and existence of well-defined hopping justify that our frequency combs map into a synthetic frequency dimension~\cite{SFanReview2018}. On the other hand, we rewrite the excited frequency combs as $A_m(\xi,\omega):=|A_m(\xi,\omega)|\exp\{\text{i}\phi_\text{S}(\xi)\}$ that are stable and can propagate up to a few nonlinear propagation length $x\approx2L_\text{NL}$. Specifically, we achieve invariant frequency combs for $|\omega|<\omega_\text{ch}$ and $-5.5L_\text{NL}<x<4L_\text{NL}$ as it is clearly shown in Fig.~\ref{Fig:Fig_Sone_SM}(a) with a well-established phase variation $\phi_\text{S}(\xi)$ but without serious amplitude distortion. Consequently, the frequency combs are spatially connected via a deterministic phase that can be achieved via Fig.~\ref{Fig:Four}(a) and (c). We consider this phase variation as a hopping between lattice sites in spatial space with hopping $v_j$.

\begin{figure}[t]
    \centering
    \includegraphics[width=0.99\textwidth]{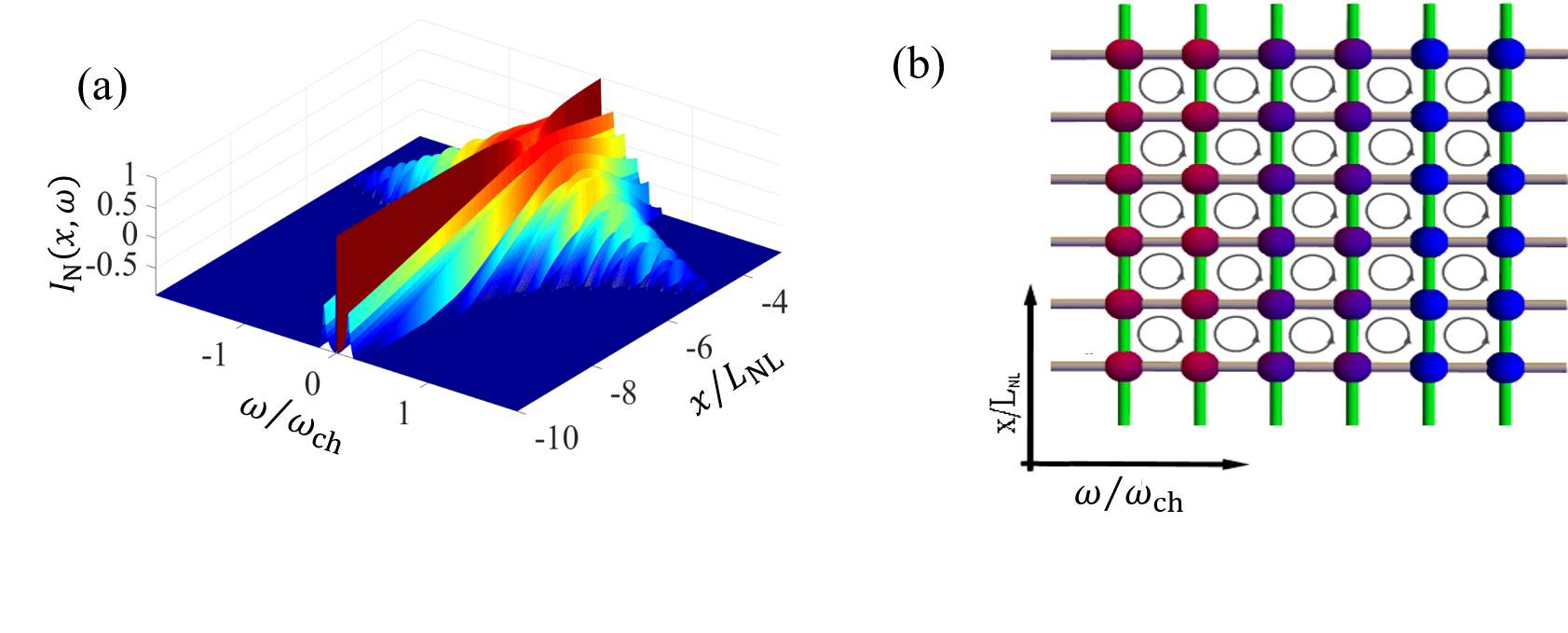}
    \caption{Spectral dynamics of localized NSPPs through nonlinear Schr\"odinger equation and its corresponding synthetic lattice: panel (a) represents the evolution of spectral harmonic side-band amplitude of the SPPs for the case of Akhmediev breather excitation for $a=0.415$, $P_{0}=10~\mu\text{W}$, $\mathcal{W}\approx3\times10^{-11}\text{s}^{2}\cdot\text{cm}^{-1}$. Other parameters of the simulation is represented in the main text. Panel (b) of this figure also denotes the two-dimensional synthetic lattice of the excited frequency combs of NSPPs in the presence of the invariants. See the text for more details.}
   \refstepcounter{SIfig} \label{Fig:Fig_Sone_SM}
\end{figure}
In our analysis, we consider the $\pi$ phase shift due to apparition of phase singularity and NSPP formation through Fermi-Pasta-Ulam-Tsingou recurrence~\cite{SPhysRevX.8.041017}. Following this assumption, we choose $N:=L_{\alpha}/L_\text{NL}$ with
\begin{equation}
    L_{\alpha}=\frac{1}{\bar{\alpha}_\text{eff}},\quad L_\text{NL}=\frac{1}{\mathcal{W}P_0},
\end{equation}
as the number of lattice sites in spatial dimension. By setting experimentally feasible parameters, the phase singularities excite for both peregrine and Akhemediev breather for $x_\mathcal{S}\approx L_\alpha/2$, $N_\mathcal{S}\approx N/2$. Therefore, we assume the phase pattern for spatial coordinate as
 \begin{equation}
    \phi_\text{S}= 
    \begin{cases}
    \phi_\text{NL}\qquad 0<j<N_\mathcal{S},\\
    \pi-\phi_\text{NL}\qquad N_\mathcal{S}<j<N,
    \end{cases}
    .
 \end{equation}
We also use $i,j$ dummy variables to represent the hopping along $\omega,x$ directions respectively. We define the hopping along $\omega$ axis as
\begin{equation}
    w_{i,j}=A_{i}(\xi_j)A_{i+1}^{*}(\xi_j),
    \label{Eq:Hopping_Spatial}
\end{equation}
for $A_{i}:=A(\xi_{i})$ characterize as spectral harmonic side-band amplitudes of NSPPs~(see Fig.~\ref{Fig:Fig_SM_Three}(a) for more details). Based on the reciprocity properties of the system we assume $A_{m,j}(\xi)=A_{-m,j}(\xi)$ and achieve the hopping for specific spatial position $\xi$ through Eq.~\eqref{Eq:Hopping_Spatial}. Moreover we assume
\begin{equation}
    v_{i,j}:=\exp\{\text{i}\varphi_{i,j}\},
\end{equation}
with
\begin{equation}
   \varphi_{i,j}:=\mathcal{K}(\omega_{i})x-\phi_\text{S},
\end{equation}
to characterize the spatial hopping for geometrical dimension, as it is clearly shown in Fig.~\ref{Fig:Fig_SM_Three}(b). Consequently, we construct the synthetic lattice with well-established hopping in both spatial and spectral dimensions that can be used to investigate the dynamical evolution of the nonlinear SPP fields.

\subsubsection{\label{Sec:Synthetic_Hamiltonian} SPL Hamiltonian and validity of lattice description}
In this section we evaluate the Hamiltonian of our synthetic lattice and achieve the dynamics of the nonlinear SPP field using this multidimensional structure. Aforementioned explanations indicate that a square lattice with two basis vector $\bm{e}_x$ and $\bm{e}_{\omega}$ describes our plasmonic frequency combs. Consequently, we describe the dynamical evolution of the frequency combs within this hybrid nonlinear interface using a well-defined two-dimensional lattice with complex hopping as it is shown in Fig.~\ref{Fig:Fig_Sone_SM}~(b).

Then, we achieve the Hamiltonian of this synthetic lattice $\mathcal{H}_\text{SL}$ as
\begin{equation}
    \mathcal{H}_\text{SPL}=\sum_{i=-\mathcal{N}}^{\mathcal{N}}\sum_{j=0}^{N}\sum_{{n}}w_{i,j}\hat{a}_{i,j}\hat{a}_{i+n,j}^{\dagger}+\sum_{i=-\mathcal{N}}^{\mathcal{N}}\sum_{j=0}^{N}v_{i,j}\hat{a}_{i,j}\hat{a}_{i,j+1}^{\dagger}+\text{H.C.},
\end{equation}
for ${n}$ the order of coupling to other harmonic side-bands in frequency dimension. We achieve this Hamiltonian through existence the invariance of energy $E$, the number of excited plasmon modes $\mathcal{N}$ and we describe the evolution of the frequency combs within this synthetic lattice through 
\begin{equation}
    \ket{\psi(x)}=\exp\{\text{i}\mathcal{H}_\text{SPL}x\}\ket{\psi(x=x_0)}.
\end{equation}
\begin{figure}[t]
    \centering
    \includegraphics[width=0.99\textwidth]{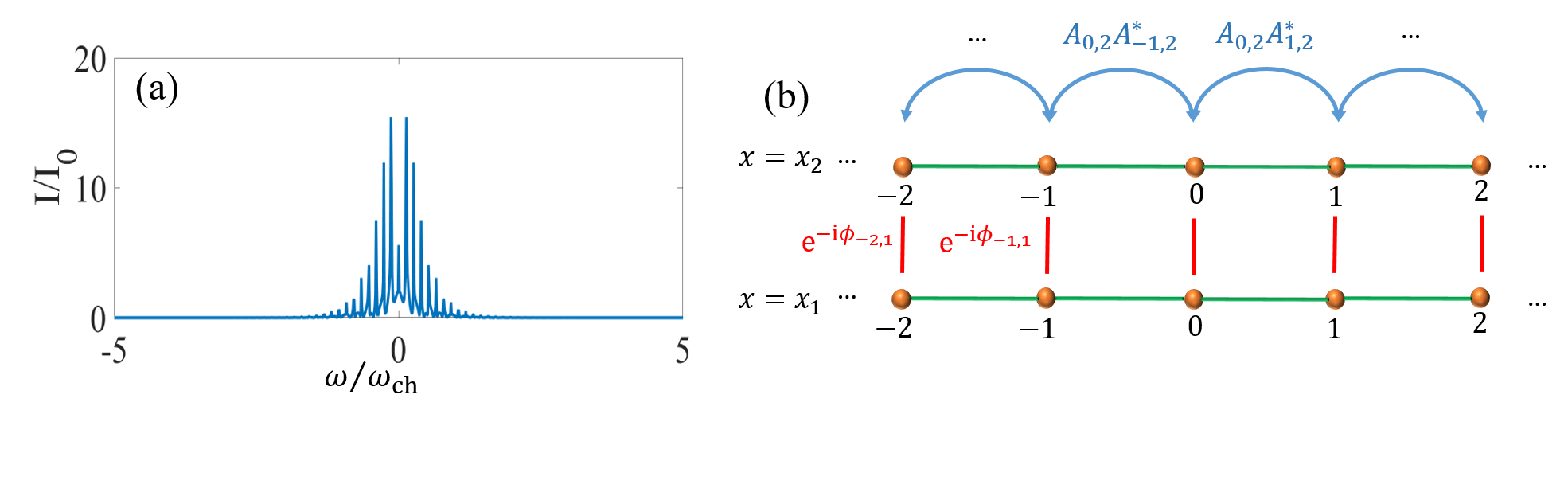}
    \caption{Explicit mapping between the nonlinear SPP frequency combs and synthetic lattice for Akhmediev breather excitation. Panel (a) is an example of the stable propagation of nonlinear SPP field as Akhemediev breather, and panel (b) is the qualitative representation of the coupling between the different synthetic lattice. The parameter used for this simulation is the same as Fig.~\ref{Fig:Fig_Sone_SM}.}
    \refstepcounter{SIfig}\label{Fig:Fig_SM_Three}
\end{figure}
In this work, we consider two kinds of NSPPs. For Akhmediev breather the spectral harmonic side-band amplitudes are obtained by Eqs.~\eqref{Eq:Zeroth_sidebandAB},~\eqref{Eq:Higher_Sideband_AB}, whereas for peregrine waves, we achieve the side-band by direct integration of Fourier transform or by calculating the limiting case of Akhmediev breather for $a\mapsto0.5$.

\end{widetext}



\begin{thebibliography}{99}
\bibitem[S23]{explanation_One}
We can derive Eq.~\eqref{Eq:Denominator} through solving Poisson equation commensurate with boundary conditions.
\end{thebibliography}
\end{document}